\documentclass[journal]{IEEEtran}
\usepackage{amssymb}
\usepackage{amsmath}
\usepackage{amsfonts}
\usepackage{empheq}
\usepackage{color}
\usepackage{cancel}

\newcommand{\beq}{\begin{equation}}
\newcommand{\eeq}{\end{equation}}
\newcommand{\bea}{\begin{eqnarray}}
\newcommand{\eea}{\end{eqnarray}}

\newcommand{\comment}[1]{}

\usepackage{cite}
\ifCLASSINFOpdf
  \usepackage[pdftex]{graphicx}
\else
  \usepackage[dvips]{graphicx}
\fi

\begin{document}
\title{Netcast: Low-Power Edge Computing with WDM-defined Optical Neural Networks}

\author{Ryan~Hamerly,~\IEEEmembership{Member,~Optica,}
		Alexander~Sludds,~\IEEEmembership{Student Member,~Optica,}
		Saumil~Bandyopadhyay,
		Zaijun~Chen,
		Zhizhen~Zhong,~\IEEEmembership{Member,~Optica,}
		Liane~Bernstein,~\IEEEmembership{Student Member,~Optica,}
%		Manya~Ghobadi,
		Dirk~Englund,~\IEEEmembership{Member,~IEEE}% As of 2021
		% <-this % stops a space
\thanks{R.H., A.S., S.B., Z.C., and D.E.\ are with MIT Research Laboratory of Electronics, Cambridge, MA 02139, USA}
\thanks{R.H.\ is with NTT Research Inc., Physics \& Informatics Laboratories, Sunnyvale, CA 94085, USA}
\thanks{Z.Z.\ %and M.G.\ are 
is
with MIT Computer Science and Artificial Intelligence Laboratory, Cambridge, MA 02139, USA}
\thanks{Manuscript received \today.}}


\markboth{Journal of \LaTeX\ Class Files,~Vol.~14, No.~8, August~2015}%
{Shell \MakeLowercase{\textit{et al.}}: Bare Demo of IEEEtran.cls for IEEE Journals}



\maketitle

\begin{abstract}
This paper analyzes the performance and energy efficiency of Netcast, a recently proposed optical neural-network architecture designed for edge computing.  Netcast performs deep neural network inference by dividing the computational task into two steps, which are split between the server and (edge) client: (1) the server employs a wavelength-multiplexed modulator array to encode the network's weights onto an optical signal in an analog time-frequency basis, and (2) the client obtains the desired matrix-vector product through modulation and time-integrated detection.  The simultaneous use of wavelength multiplexing, broadband modulation, and integration detection allows large neural networks to be run at the client by effectively pushing the energy and memory requirements back to the server.  The performance and energy efficiency are fundamentally limited by crosstalk and detector noise, respectively.  We derive analytic expressions for these limits and perform numerical simulations to verify these bounds.
\end{abstract}


% \fi
\IEEEpeerreviewmaketitle

\section{Introduction}
\IEEEPARstart{M}{achine} learning is widespread in the cloud, where ample processing power is co-located with data, but in recent years, network and privacy constraints have pushed processing closer to the end user \cite{Yu2017}.  This ``edge computing'' paradigm introduces new constraints to hardware and software design, as data processing happens on size, weight and power (SWaP)-constrained devices at the edge of the network.  In light of the growing importance of deep neural networks (DNNs) in information processing, significant effort has been dedicated to SWaP-constrained hardware \cite{Sze2017, Chen2014} and algorithms \cite{Howard2017, Iandola2016} for DNN edge inference.  While these efforts have enabled edge deployment of intermediate-size neural networks \cite{Xu2018}, many state-of-the-art DNNs are still too large \cite{Szegedy2016, Dai2019} to be efficiently run on the edge.

Consequently, edge computing is a prime target for emerging hardware technologies.  Memristive circuits \cite{Krestinskaya2019} are a leading candidate, leveraging the standard electronics process as well as the substantial body of work towards edge inference on DNN crossbar arrays.  However, practical issues relating to uniformity, accuracy, and resistance updates remain outstanding challenges for the field \cite{Ambrogio2018}; moreover, the weight stationary \cite{Sze2017} character of memristor networks means SWaP constraints will still set an upper limit on the DNN size, and will limit the frequency of network updates.  Photonic architectures have also gained traction, owing to the physical mapping between matrix-vector multiplication (the hardware bottleneck in DNN inference \cite{Sze2017}) and linear-optical processes.  However, scaling issues are much more pronounced in photonics, and usually involve a tradeoff between speed, size, and programmability \cite{Paek1987, New2017, Shen2017, Prabhu2020, Tait2017, Pai2022, Bernstein2022}.  Moreover, in integrated architectures, chip area constraints, loss, and error propagation \cite{Fang2019, SaumilPaper, RyanPaper1, RyanPaper2, RyanPaper3, Alexiev2021} make large networks especially challenging to implement.

Recently, we introduced Netcast, an optical neural-network architecture that combines the advantages of wavelength division multiplexing (WDM), broadband modulation, and integration detection \cite{Netcast1, Sludds2022, Zhong2021}.  Our protocol consists of two components: a WDM modulator array (server), which is connected by an optical link to the SWaP-constrained edge device (client) which consists of a single modulator, a demultiplexer, and a set of time-integrating detectors.  For each DNN layer, over a sequence of time steps, the server encodes the weight matrix as an analog optical waveform, with matrix elements stored in a time-frequency basis.  At the client, this signal is modulated and demultiplexed, and the charge accumulated on the detectors encodes the neuron activations for the next DNN layer.  Netcast leverages the high bandwidth of WDM optical links to effectively ``split'' the computation into two parts, pushing the hard part of the computation to the server while the client only performs a minimal postprocessing step.  This enables low-power DNN edge inference for networks of arbitrary size, unbounded by either power or memory constraints of edge devices.  In this paper, we analyze the limits to the performance and energy consumption of Netcast, which are set by crosstalk and detector noise, respectively.  We derive analytic expressions for these bounds, which are verified with numerical simulations.

\section{Netcast Concept}

\begin{figure}[htbp]
\begin{center}
\includegraphics[width=1.00\columnwidth]{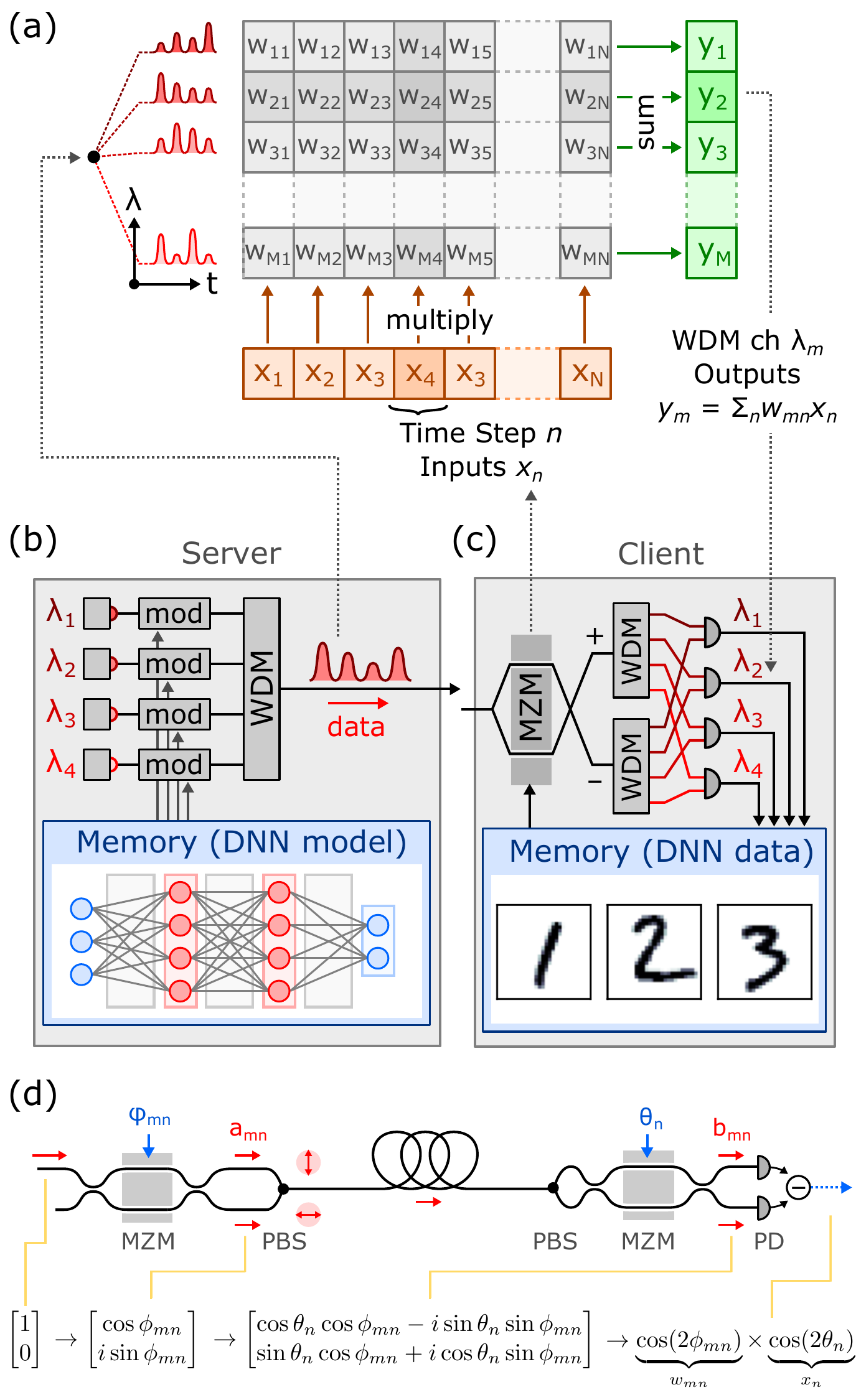}
\caption{Schematic of Netcast architecture.  (a) Ouput-stationary dataflow for MVM.  At each time step, a column $w_{:,n}$ is scaled by $x_n$ and accumulated to the partial sums for $y_m$.  (b) Weight server consisting of a WDM bank of optical modulators and a large memory containing the full DNN model.  (c) Client consisting of a single broadband modulator (e.g.\ balanced MZM), demultiplexing optics, and integrating detectors.  (d) Optical dataflow for a single wavelength channel in the specific case where modulation is performed with balanced MZMs.}
\label{fig:f1}
\end{center}
\end{figure}

\label{sec:concept}

Fig.~\ref{fig:f1} illustrates the concept.  The architecture consists of a {\it server} and a {\it client}, connected by an optical {\it link}.  Since linear algebra is the bottleneck step for DNN inference and training, we focus on how the optical hardware accelerates matrix-vector multiplication (MVM) $y_m = \sum_n{w_{mn}x_n}$ at the client; the nonlinear activation function, pooling, and batch normalization can be performed at the client with minimal added cost.  In the output-stationary dataflow \cite{Sze2017}, an $M\times N$ MVM is performed in $N$ time steps, where in each step a column of $w$ is weighted by an element of $x$ and accumulated to the partial sums for $y$, i.e.\  \texttt{y[m] += w[m,n]*x[n]} (Fig.~\ref{fig:f1}(a)).  To perform this operation optically, weight data is encoded in an $N$-step pulse train over $M$ wavelength channels, with the rows and columns of the matrix represented by time and wavelength, respectively.  This optical signal can be generated using a WDM modulator bank as shown in Fig.~\ref{fig:f1}(b).  At each time step (indexed by $n$), the client receives a column $w_{:,n}$ of the weight matrix, and modulates this signal using a broadband modulator in order to scale the column by $x_n$.  The resulting output is demultiplexed into an array of $M$ difference photodetectors (one per wavelength) as shown in Fig.~\ref{fig:f1}(c), which produce the products $w_{mn}x_n$.  Integrating over all $N$ time steps, the charge accumulated on the detectors yields the MVM output.

To understand this scheme quantitatively, Fig.~\ref{fig:f1}(d) traces the protocol in more detail.  Here, we consider only the path of a single wavelength channel $\lambda_m$, since the channels process data independently (but see Sec.~\ref{sec:tput}). For concreteness, here we employ a balanced Mach-Zehnder modulator (MZM) at both the server and client.  The server-side MZM (followed by a 90$^{\rm o}$ phase shift) splits the input into two channels with amplitudes $\vec{a}_{mn} \equiv (a_{mn}^{(+)}, a_{mn}^{(-)})$, given by:
\beq
	\vec{a}_{mn} %\equiv \begin{bmatrix} a_{mn}^{(+)} \\ a_{mn}^{(-)} \end{bmatrix} 
	\propto \begin{bmatrix} 1 & 0 \\ 0 & -i \end{bmatrix}
	\underbrace{\begin{bmatrix} \cos\phi_{mn} & i\sin\phi_{mn} \\ i\sin\phi_{mn} & \cos\phi_{mn} \end{bmatrix}}_{T(\phi_{mn})} \begin{bmatrix} 1 \\ 0 \end{bmatrix}
	= \begin{bmatrix} \cos\phi_{mn} \\ \sin\phi_{mn} \end{bmatrix} \label{eq:mzm1}
\eeq
which encode the weight through differential signaling $w_{mn} = |a_{mn}^{(+)}|^2 - |a_{mn}^{(-)}|^2$ (analogous to the weight banks of Ref.~\cite{Tait2017}).  Next, a polarization beamsplitter (PBS) combines the through- and drop-port outputs to the orthogonal polarizations of the optical link, where the signal is transmitted to the client.  Links may be over fiber or free space, and may include optical fan-out to multiple clients.  If the link loss or fan-out ratio is large, the server output can be pre-amplified.

%\begin{figure}[tbp]
%\begin{center}
%\includegraphics[width=1.0\textwidth]{Note15-Fig2.pdf}
%\caption{Data flow in the {\sc NetCast} ONN.  A matrix-vector product is performed in $N$ time steps, with $M$ wavelength channels.  In each time step $n$, the weights $w_{mn}$ are encoded by adjusting the detunings $\Delta_{mn}$ of the rings.  The through- and drop-port outputs $a^{\rm (T)}_{mn} = t_{mn} a_0$, $a^{\rm (D)}_{mn} = r_{mn} a_0$ (Eq.~(\ref{eq:tr})) are sent to the client, where the MZM mixes them to produce outputs $a_{mn}^{(\pm)}$ (Eq.~(\ref{eq:mzm})).  The difference current in each wavelength channel gives the product $w_{mn}x_n$.  After time integration, the products $y_m = \sum_{n}w_{mn} x_n$ are read out.}
%\label{fig:f2}
%\end{center}
%\end{figure}

At the end of the link, the signal enters the client (Fig.~\ref{fig:f1}(d), right), where a second PBS (after any necessary polarization correction, not shown) separates the polarizations and a phase shifter is used to correct for any relative phase shift accrued in the link, effectively recovering the signal pair $\vec{a}_{mn}$ from before the server-side PBS.  These inputs are scrambled using a two-port broadband MZM, whose voltage encodes the current activation $x_n$.  At the output of this MZM we have:
\beq
	\vec{b}_{mn} = T(\theta_n) %\begin{bmatrix} \cos\theta_n & i\sin\theta_n \\ i\sin\theta_n & \cos\theta_n \end{bmatrix} 
	\vec{a}_{mn}
	= \begin{bmatrix} \cos\theta_n \cos\phi_{mn} + i\sin\theta_n\sin\phi_{mn} \\
	\cos\theta_n\sin\phi_{mn} + i\sin\theta_n\cos\phi_{mn} \end{bmatrix}
	\label{eq:mzm}
\eeq
Finally, the WDM channels are demultiplexed and the power in each channel is read out on a photodetector.  We care about the difference current between the MZM outputs, integrated over $N$ time steps, which evaluates to:
\begin{align}
	& Q_{m} \propto \sum_n \bigl[|b_{mn}^{(+)}|^2 \!-\! |b_{mn}^{(-)}|^2\bigr]
	\propto \sum_n \underbrace{\cos(2\phi_{mn})}_{\text{input}\,w_{mn}} \times  \underbrace{\cos(2\theta_n)}_{\text{input}\,x_n} \nonumber \\
	\label{eq:di}
\end{align}
In this way, for inputs and weights scaled to the range $x_n, w_{mn} \in [-1, +1]$, and with the encodings $\phi_{mn} = \tfrac12 \cos^{-1}(w_{mn})$ and $\theta_n = \tfrac12 \cos^{-1}(x_{n})$, the client will generate a signal proportional to $y_m$, performing the desired the matrix-vector product.

While Eq.~(\ref{eq:di}) is specific to MZMs, the Netcast architecture is compatible with a range of modulators, subject to the constraint that the client-side modulator be optically broadband.  For example, micro-ring resonators (MRRs) are a particularly compact, scalable, and low-energy (server-side) modulator type \cite{Timurdogan2014, Stojanovic2018, Haffner2018, DeCea2021, Rizzo2021}.  While MRRs are often deemed unsuitable for high-precision modulation, recent demonstrations of MRR control up to 9~bits of precision \cite{Zhang2022} highlight the potential for this platform.  In a two-port MRR critically coupled to the input port (losses $\kappa_1 = \kappa_2 + \kappa_{\rm abs}$, programmable detuning $\Delta_{mn}$), the same analysis leading to Eqs.~(\ref{eq:mzm1}-\ref{eq:di}) yields $Q_m \propto (\Delta_{mn}-\kappa_1\kappa_2)/(\Delta_{mn}-\kappa_1^2) \times \cos(2\theta_n)$.  If weights in the range $w_{mn} \in [-\kappa_2/\kappa_1, +1]$ are encoded into the MRR detuning as $\Delta_{mn} = \kappa_1 \sqrt{(\kappa_2/\kappa_1 + w_{mn})/(1-w_{mn})}$, then the charge $Q_m$ will evaluate to the desired matrix-vector product.

The main insight underlying Netcast is not that photonics provides a means to add and multiply numbers, but that we can (1) perform many operations in parallel and (2) separate the tasks of logic (client) and memory access (server) by means of an optical link.  Although the server and client cooperate to perform the calculation, the workloads are unequal, and for large DNNs, the energy and memory costs at the client are significantly lower than those at the server:
\begin{itemize}
	\item For a layer of size $N\times N$, the memory cost scales as $O(N^2)$ at the server and $O(N)$ at the client.  In general, memory reads (especially from off-chip DRAM) dominate the energy consumption in DNN hardware \cite{Sze2017}.  Additionally, the server must drive $N$ modulators for a total energy cost of $O(N^2)$, while the client only drives a single modulator.  The client reads out $N$ detectors, but only after integration ($O(N)$ cost).
	\item The server must store the full neural network, giving a memory requirement of $O(N^2L)$, where $L$ is the number of layers; by contrast, the client stores only the activations, which requires a much smaller memory of size $O(N)$.  This discrepancy is consistent with the general observation in neuroscience that the ratio of synapses to neurons is very large. %\note{ref}.
	\item Although the client-side energy cost only scales as $O(N)$, the client performs the full $N^2$ operations needed for an MVM.  The broad optical bandwidth of MZMs%\note{ref}
	, coupled with the use of integrating detectors \cite{Hamerly2019}, allows the client to leverage an $O(N)$ optical parallelism factor \cite{Spectrum}, reducing the client-side MVM cost from quadratic to linear scaling.
\end{itemize}

By means of a single-mode optical link, the Netcast scheme pushes all of the difficult parts of the computation to the server, liberating the edge client from a significant part of its SWaP constraints.  This can in principle enable the edge deployment of entirely new classes of DNNs that have up to now been restricted to use in data centers.

\begin{figure}[b!]
\begin{center}
\includegraphics[width=1.00\columnwidth]{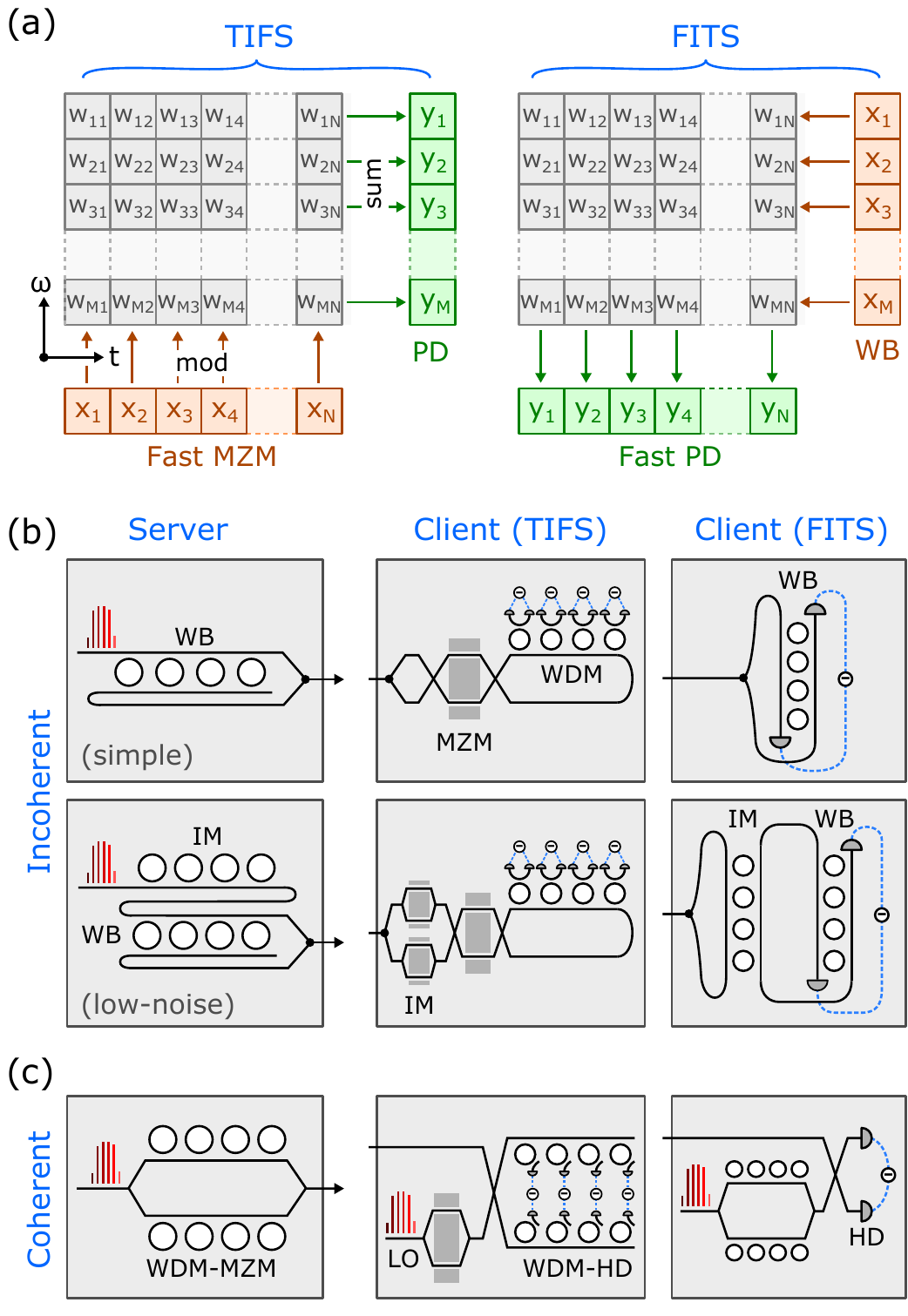}
\caption{Variants of optical implementation.  (a) MVM implemented by time integration / frequency separation (left) and frequency integration / time separation (right).  (b) Server and client architectures for incoherent encoding.  Here a WDM ring array is depicted for the server modulators.  One can mix and match server and client designs.  (c) Coherent encoding architectures.}
\label{fig:f2}
\end{center}
\end{figure}

\section{Variations}
\label{sec:var}

The key concepts of Netcast---WDM, parallel modulation, and integration detection---admit a number of variations, the most obvious ones depicted in Fig.~\ref{fig:f2}.  All of these schemes encode the weight matrix in time-frequency space, where $w_{mn}$ is the amplitude of wavelength band $\lambda_m$ at time step $t_n$.  Two matrix-vector multiplications are possible (Fig.~\ref{fig:f2}(a)): right-multiplication $y = w x$ through Time Integration / Frequency Separation (TIFS), or left-multiplication $y^T = x^T w$ through Frequency Integration / Time Separation (FITS).  The system presented in Fig.~\ref{fig:f1} used TIFS.  For FITS, the client uses a weight bank (WB) \cite{Tait2017} consisting of an array of ring resonators, which integrates over frequency with the activations $x_m$ encoded in the resonator detunings (Fig.~\ref{fig:f2}(b)).  Time separation requires a single fast detector pair, unlike the TIFS schemes where $M$ slow detectors are used.

Another distinction is the modulation format.  The approach in Fig.~\ref{fig:f1} shows the simplest case: incoherent differential signaling with fixed power.  However, for small differential signals, shot noise can present a challenge to this approach, but with additional hardware complexity at either the server or client, one can lower this receiver noise (Sec.~\ref{sec:ln}).  Additionally, with a client-side local oscillator, one can encode the signal coherently, which increases bandwidth and reduces noise to the quantum limit (Sec.~\ref{sec:coh}).  Altogether, this leads to $2^2\times 2 + 1^2 \times 2 = 10$ possibilities, described below.

\subsection{Differential Noise Reduction} 
\label{sec:ln}

\begin{table*}[t!]
\begin{center}
\caption{Comparison of the four incoherent schemes and the coherent scheme.  $^\dagger$Weight and PD input powers for case $w_{mn} > 0$, $x_n > 0$ shown.  The other cases are analogous and the forms for $Q_{\rm tot}$, $Q_{\rm det}$ are the same.}
\begin{tabular}{r@{/}l|c|c|c|cc|cc}
\hline\hline
\multicolumn{2}{c|}{\bf Scheme} & \multicolumn{2}{c|}{{\bf Transmitter}$^{\dagger}$} & \multicolumn{3}{c|}{{\bf Detector}$^{\dagger}$} & \multicolumn{2}{c}{{\bf Noise}} \\
\multicolumn{2}{c|}{} & \multicolumn{1}{c}{$|a_\pm|^2/N_{\rm src}$} & $N_{\rm tr}/N_{\rm src}$ & \multicolumn{1}{c}{$|b_\pm|^2/N_{\rm src}$} & $Q_{\rm det}/N_{\rm src}$ & \!\!$Q_{\rm tot}/N_{\rm src}$\!\! & $F_{\rm src}$ & $F_{\rm tr}$ \\ 
\hline
S&S & $\tfrac{1}{2}(1 \pm w_{mn})$ & 1 & $\tfrac{1}{2}(1 \pm w_{mn}x_n)$ & $w_{mn}x_n$ & $1$ 
& 1 & 1 \\
S&LN & $\tfrac{1}{2}(1 \pm w_{mn})$ & 1 & $\tfrac{1}{2}(1 \pm w_{mn})x_n$ & $w_{mn}x_n$ & $|x_n|$ 
& $\langle |x_n| \rangle$ & $\langle |x_n| \rangle$ \\
LN&S & $\{w_{mn},\ 0\}$ & \!\!$\langle|w_{mn}|\rangle$\!\! & $\tfrac{1}{2}w_{mn} (1 \pm x_{n})$ & $w_{mn}x_n$ & $|w_{mn}|$ 
& $\langle |w_{mn}| \rangle$ & $\langle |w_{mn}| \rangle^2$ \\
LN&LN & $\{w_{mn},\ 0\}$ & \!\!$\langle|w_{mn}|\rangle$\!\! & $\{w_{mn}x_n,\ 0\}$ & $w_{mn}x_n$ & \!$|w_{mn}x_n|$\! 
& \!\!$\langle |w_{mn}x_n| \rangle$\!\! & \!\!\!$\langle |w_{mn}| \rangle \langle |w_{mn}x_n| \rangle$\!\!\! \\ 
\hline
\multicolumn{2}{c|}{Coherent\!} & \!\!$a_{\rm wt} = w_{mn} \alpha_{\rm src}$\!\! & \!\!$\langle|w_{mn}|^2\rangle$\!\! & \!\!$\tfrac{1}{2}(\alpha_x x_n \pm \alpha_{\rm src} w_{mn})^2\vphantom{\Bigr|}$\!\! & \!\!$2\alpha_x\alpha_{\rm src} w_{mn}x_n$\!\! & $\alpha_x^2 |x_n|^2$ 
& $\tfrac{1}{4}\langle |x_n|^2 \rangle$ & \!$\tfrac{1}{4}\langle |w_{mn}|^2 \rangle\langle |x_n|^2 \rangle$\!
\\
\hline\hline
\end{tabular}
%\begin{tabular}{r@{/}l|c|c|c|cc}
%\hline\hline
%\multicolumn{2}{c|}{\bf Scheme} & \multicolumn{2}{c|}{{\bf Transmitter}$^{\dagger}$} & \multicolumn{3}{c}{{\bf Detector}$^{\dagger}$}  \\
%\multicolumn{2}{c|}{} & \multicolumn{1}{c}{$|a_\pm|^2/N_{\rm src}$} & $N_{\rm tr}/N_{\rm src}$ & \multicolumn{1}{c}{$|b_\pm|^2/N_{\rm src}$} & $Q_{\rm det}/N_{\rm src}$ & \!\!$Q_{\rm tot}/N_{\rm src}$\!\! \\
%\hline
%S&S & $\tfrac{1}{2}(1 \pm w_{mn})$ & 1 & $\tfrac{1}{2}(1 \pm w_{mn}x_n)$ & $w_{mn}x_n$ & $1$ \\
%S&LN & $\tfrac{1}{2}(1 \pm w_{mn})$ & 1 & $\tfrac{1}{2}(1 \pm w_{mn})x_n$ & $w_{mn}x_n$ & $|x_n|$ \\
%LN&S & $\{w_{mn},\ 0\}$ & \!\!$\langle|w_{mn}|\rangle$\!\! & $\tfrac{1}{2}w_{mn} (1 \pm x_{n})$ & $w_{mn}x_n$ & $|w_{mn}|$  \\
%LN&LN & $\{w_{mn},\ 0\}$ & \!\!$\langle|w_{mn}|\rangle$\!\! & $\{w_{mn}x_n,\ 0\}$ & $w_{mn}x_n$ & \!$|w_{mn}x_n|$\! 
% \\ 
%\hline
%\multicolumn{2}{c|}{Coherent\!} & \!\!$a_{\rm wt} = w_{mn} \alpha_{w}$\!\! & \!\!$\langle|w_{mn}|^2\rangle$\!\! & \!\!$\tfrac{1}{2}(\alpha_x x_n \pm \alpha_{w} w_{mn})^2\vphantom{\Bigr|}$\!\! & \!\!$2\alpha_x\alpha_{w} w_{mn}x_n$\!\! & $\alpha_x^2 |x_n|^2$ 
%\\
%\hline\hline
%\end{tabular}
\label{tab:t1}
\end{center}
\end{table*}

As noted, a critical challenge to the simple differential detection scheme of Fig.~\ref{fig:f1}(d) is the poor signal-to-noise ratio for small inputs, since the shot noise is proportional to the total charge $Q_{\rm tot} \propto \bigl[|b_{mn}^{(+)}|^2 + |b_{mn}^{(-)}|^2\bigr]$, which is a constant.  Therefore, to resolve a signal of order $\epsilon \ll 1$, the receiver requires $O(1/\epsilon^2)$ photons to beat the shot noise.  In realistic DNNs, many weights are small or pruned to zero, so in order to obtain a reasonable signal-to-noise ratio, unacceptably high optical powers may be required.  To circumvent this difficulty, one can prepend an intensity modulator to the server or client (Fig.~\ref{fig:f2}(b)), creating a ``low-noise'' device where differential signaling on small signals uses small total optical powers, substantially reducing the shot noise.

To show the advantage of the ``low-noise'' configurations, here we consider the four cases, named S/S, S/LN, LN/S, LN/LN (simple server / simple client, etc.).  In all cases, we start with an unweighted WDM light source with amplitudes $\alpha_w$, where $N_{\rm src} = |\alpha_w|^2$ is the number of photons per weight (at the source), and normalize variables so that $w, x \in [-1, 1]$.
\begin{enumerate}
	\item {\it S/S}: The weight bank (WB) encodes $w_{mn}$ into the differential power in two channels, multiplexed with a PBS (Fig.~\ref{fig:f2}(b), top).  These are $|a_\pm|^2 = \tfrac{1}{2}(1 \pm w_{mn})N_{\rm src}$.  At the client these channels are mixed with the MZM to give $|b_\pm|^2 = \tfrac{1}{2}(1 \pm w_{mn}x_n)N_{\rm src}$.  Thus the differential charge is $Q_{\rm det} = |b_+|^2 - |b_-|^2 = w_{mn}x_n N_{\rm src}$, while the total absorbed charge, which sets the shot noise, is $Q_{\rm tot} = |b_+|^2 + |b_-|^2 = N_{\rm src}$.  
	\item {\it S/LN}: Here we use the same inputs as S/S, but the client has an additional pair of intensity modulators (IM) before the MZM (Fig.~\ref{fig:f2}(b), lower center / right).  The IMs attenuate the power according to the amplitude $|x_n|$, while the MZM works in binary mode to encode the sign ($\theta_n = \mbox{arg}(x_n) \in \{0, \pi/2\}$).  Thus the PD input is either  $|b_\pm|^2 = \tfrac{1}{2}(1 \pm w_{mn})|x_n| N_{\rm src}$ for $x_n > 0$, or $\tfrac{1}{2}(1 \mp w_{mn})|x_n| N_{\rm src}$ for $x_n < 0$.  $Q_{\rm det}$ is the same, but $Q_{\rm tot}$ is reduced by a factor of $|x_n|$.
	\item {\it LN/S}: In this case, a standard client is used but the weight server has an additional IM before the WB (Fig.~\ref{fig:f2}(b), lower left).  Like the S/LN case, the IM encodes the amplitude $|w_{mn}|$ while the WB functions in binary mode to encode the sign.  Thus, only a single polarization carries power: $a_+ = |w_{mn}| N_{\rm src}$ if $w_{mn} > 0$, and $a_- = |w_{mn}| N_{\rm src}$ if $w_{mn} < 0$.  The PD input is $|b_{\pm}|^2 = \tfrac{1}{2}|w_{mn}|\bigl(1\pm \text{sign}(w_{mn})x_n\bigr) N_{\rm src}$, which gives the same $Q_{\rm det}$, but $Q_{\rm tot}$ is reduced by a factor of $|w_{mn}|$ compared to the S/S case.
	\item {\it LN/LN}: If both server and client use the low-noise designs, all the power ends up in one of the detectors.  This is the most efficient case: either $|b_+|^2 = |w_{mn}x_n| N_{\rm src}$ for the case $w_{mn}x_n > 0$, or $|b_-|^2 = |w_{mn}x_n| N_{\rm src}$ for the case $w_{mn}x_n < 0$.  Thus $Q_{\rm tot}$ is reduced by a factor $|w_{mn}x_n|$.
\end{enumerate} 

These cases are enumerated in Table~\ref{tab:t1}.  Here we list the transmitter-side amplitudes $a_{\pm}$ and the corresponding power fraction $P_{\rm tr} = (|a_+|^2+|a_-|^2)/N_{\rm src}$, as well as the detector-side amplitudes $b_{\pm}$, the measured differential charge $Q_{\rm det} = |b_+|^2-|b_-|^2$, and the total charge $Q_{\rm tot} = |b_+|^2+|b_-|^2$.  The important point is that, while they collect the same differential charge $Q_{\rm det} = w_{mn}x_n N_{\rm src}$, the total PD charge $Q_{\rm tot}$, which sets the shot-noise limit, can be reduced considerably if many of the inputs or weights are small (or zero).

\subsection{Coherent Detection}
\label{sec:coh}

By analogy to Ref.~\cite{Hamerly2019}, one can also construct a coherent version of Netcast, depicted in Fig.~\ref{fig:f2}(c).  As in the original scheme (Fig.~\ref{fig:f1}), the weights $w_{mn}$ are generated at a server in a time-frequency basis by modulating the lines of a WDM source and broadcast to the client over an optical link.  Here we encode data in the complex amplitude of the field rather than its power, and only use a single polarization.  To measure this amplitude, we need an identical local oscillator (LO) at the client side.  Because of the large number of wavelength channels, a frequency comb may be the ideal source at both server and client; comb locking remains an active topic in telecommunications research \cite{Geng2022, Kemal2016, Jang2018, Liao2019}.  A fraction of the LO power (not shown) is mixed with the signal to generate a beatnote in order to lock the LO comb to the server.  The remainder is amplitude modulated in an MZM, which scales the LO amplitude by the activations $x_n$.  A wavelength-demultiplexed homodyne detector accumulates the products $w_{mn} x_n$, which integrate out to give the matrix-vector product just as in the incoherent case.

One advantage of the coherent scheme is bandwidth: with IQ modulation, and polarization multiplexing, one can achieve a $4\times$ higher data rate than the incoherent scheme.  In addition, the coherent scheme benefits from increased SNR, especially at low signal powers.  This is especially relevant for long-distance free-space links where the transmission efficiency is very low.  Homodyne detection with a sufficiently strong LO allows this signal to be measured down to the quantum limit, rather than being swamped by Johnson noise.

Given $x_n, w_{mn} \in [-1, 1]$ as before, the comb line amplitudes input to the homodyne detector, normalized to photon number, are $a_w = \alpha_w w_{mn}$ and $a^x = \alpha_x x_{n}$.  In the weak-signal limit $\alpha_w \ll \alpha_x$, the differential and total charge accumulated the detector pair, per time step, is (Table~\ref{tab:t1}):
\beq
	Q_{\rm det} = 2 \alpha_w \alpha_x w_{mn} x_n,\ \ \ 
	Q_{\rm tot} = \alpha_x^2 |x_n|^2
\eeq

%\note{Mention FITS can use passive WDM and EAMs, connect to Pleros work.}

\section{Energy Consumption and Noise}

Since the total number of operations scales as $O(N^2)$, when running on a single client, NetCast does not yield an improvement in {\it total} energy efficiency compared to running the DNN digitally.  However, the client-side power consumption is reduced significantly---since as discussed previously, the electrical energy consumption for an $N\times N$ layer scales as $O(N)$, which translates to an energy per MAC scaling of $O(N^{-1})$.  Figures based on current technology ($O(1)$~pJ/sample for modulation, DAC, and ADC \cite{Hamerly2019, Miller2012, Atabaki2018, Jonsson2011, Cosemans2019}, see also discussions in \cite{Sludds2022, Bernstein2022, Tait2022, Cole2021}), suggest fJ/MAC (client side) performance with matrix sizes $N \geq 100$, which is several orders of magnitude below the current (system-level) CMOS state of the art \cite{Horowitz2014, Jouppi2017, Reuther2019}.

One should also consider the optical energy consumption, as this is often tied to fundamental limits on the performance of photonic hardware \cite{Hamerly2019, Garg2021, Tait2022}.  The optical power sets the signal-to-noise ratio of the client's detectors, with the measured photocurrent taking the form:
\beq
	Q_m = \sum_n Q_{{\rm det},mn} + N(0, \sigma_Q^2),\ \ \ 
	\sigma_Q^2 = \underbrace{\vphantom{\sum_n}\frac{kTC}{e^2}}_{\text{Johnson}} + \underbrace{\vphantom{\frac{kTC}{e}} \sum_n Q_{{\rm tot},mn}}_{\text{shot}}
\eeq
The measured matrix-vector product $y_m$ is proportional to $Q$; thus noise manifests as a Gaussian error term in the analog matrix-vector multiplication
\beq
	y_m = \sum_n w_{mn} x_n + N(0, \sigma^2),\ \ \ \sigma = \sqrt{\sigma_J^2 + \sigma_S^2}
\eeq
where the terms $\sigma_J$ and $\sigma_S$ correspond to Johnson (kTC) and shot noise.  Both terms depend on the optical power, which can be defined in two ways: (1) in terms of the source power before modulation $N_{\rm src}$, or (2) in terms of the transmitted power after modulation $N_{\rm tr}$.  For the simple differential-signaling transmitted, $N_{\rm tr} = N_{\rm src}$, but for the low-noise and coherent designs, it can be much smaller for networks with many small weights (see Table~\ref{tab:t1}).  Since the server's optical output can be amplified before transmission, $N_{\rm tr}$ is likely the more practically relevant energy metric.  In terms of $N_{\rm src}, N_{\rm tr}$, the noise amplitudes scale as:
\beq
	\sigma_J^2 = \frac{kTC/e^2}{N_{\rm src}^2},\ \ \ 
	\sigma_S^2 = F_{\rm src} \frac{N}{N_{\rm src}} = F_{\rm tr} \frac{N}{N_{\rm tr}} \label{eq:sig}
\eeq
The Johnson noise term is self-explanatory.  In the shot noise term, we observe an additional constant of proportionality ($F_{\rm src}$, $F_{\rm tr}$, see Table~\ref{tab:t1}), which depends on the transmitter and detector design, since different designs have different $Q_{\rm tot}$, which sets the shot noise.  These dimensionless constants quantify the noise reduction of the coherent and low-noise schemes relative to the simple schemes and are $\ll 1$ if most of the inputs or weights are small.  %Table~\ref{tab:t1} lists the forms for $F_{\rm src}$, and $F_{\rm tr}$ used in Eqs.~(\ref{eq:sig}) for all five schemes discussed in Sec.~\ref{sec:var}.

\begin{figure}[tbp]
\begin{center}
\includegraphics[width=1.00\columnwidth]{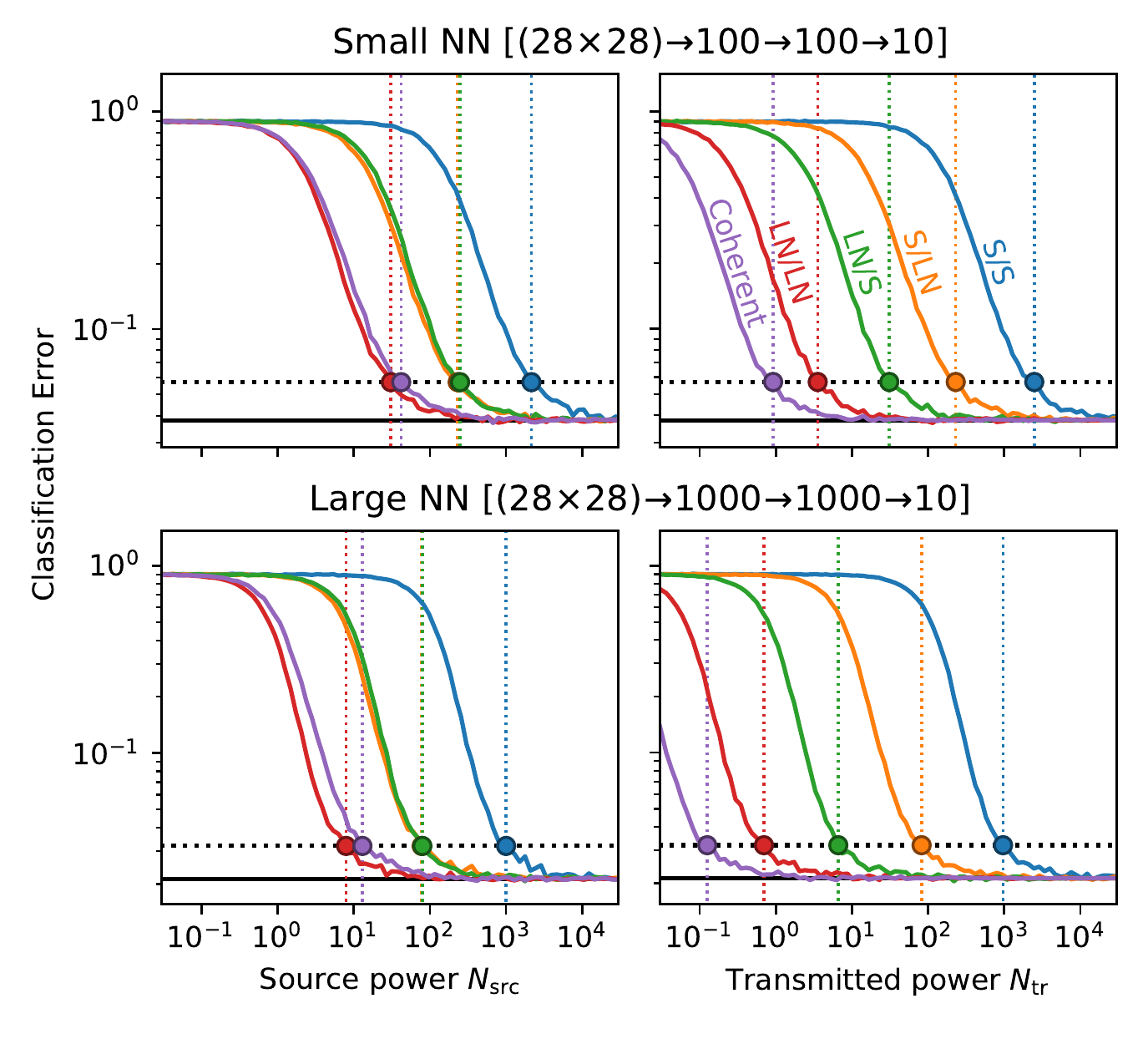}
\caption{Effect of shot noise on accuracy of MNIST fully-connected DNNs.  Left column: dependence on $N_{\rm src}$ (same value used for all layers).  Right column: dependence on $N_{\rm tr}$.  Circles denote the point at which the DNN accuracy is degraded by a 50\% error rate increase.}
\label{fig:f3}
\end{center}
\end{figure}

\begin{figure}[tbp]
\begin{center}
\includegraphics[width=1.00\columnwidth]{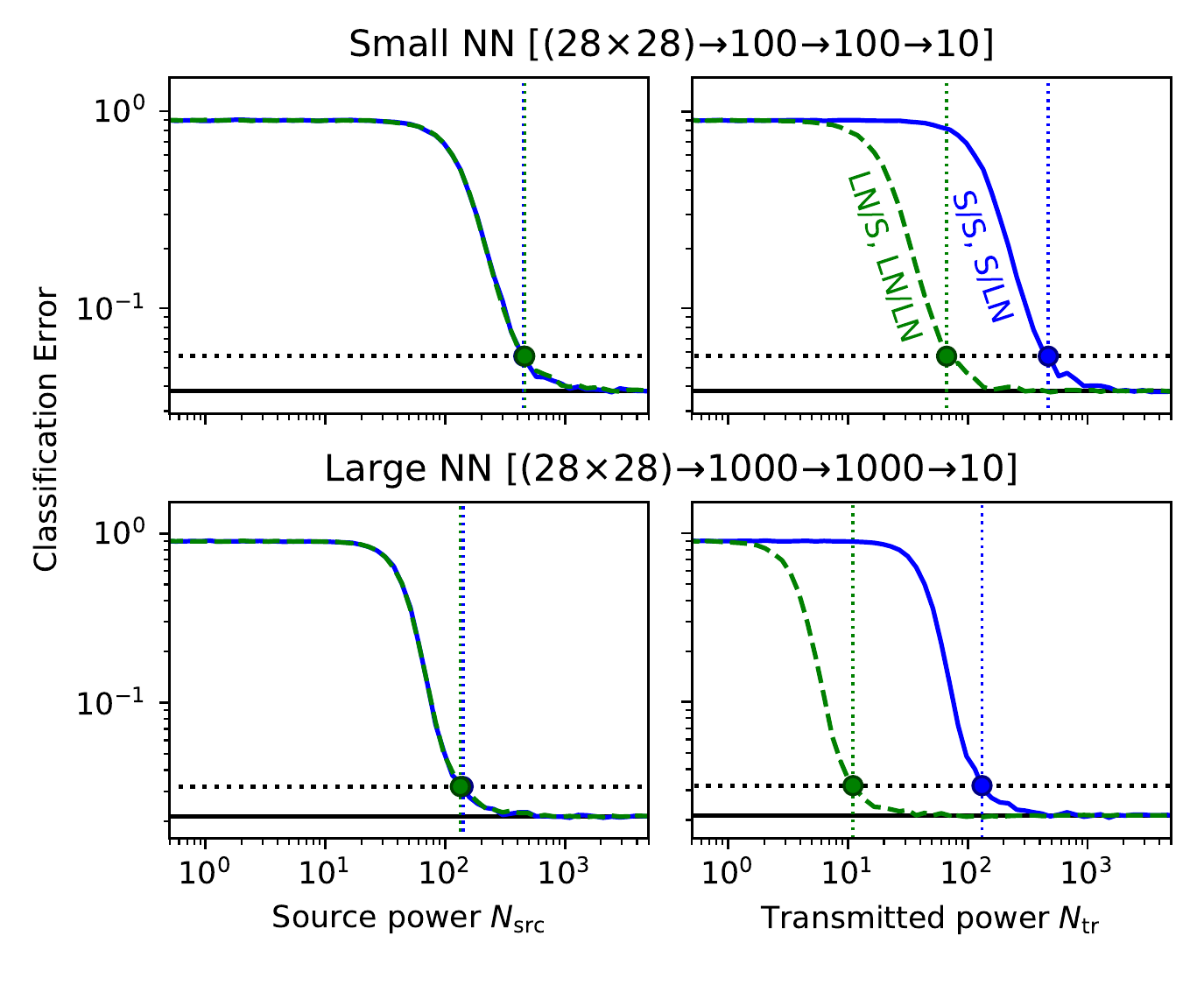}
\caption{Effect of Johnson noise on DNN accuracy ($C = 0.1$~pF).  Since $\sigma_J$ depends only on $N_{\rm src}$, the curves in the left column coincide.}
\label{fig:f4}
\end{center}
\end{figure}

\begin{figure}[tbp]
\begin{center}
\includegraphics[width=1.00\columnwidth]{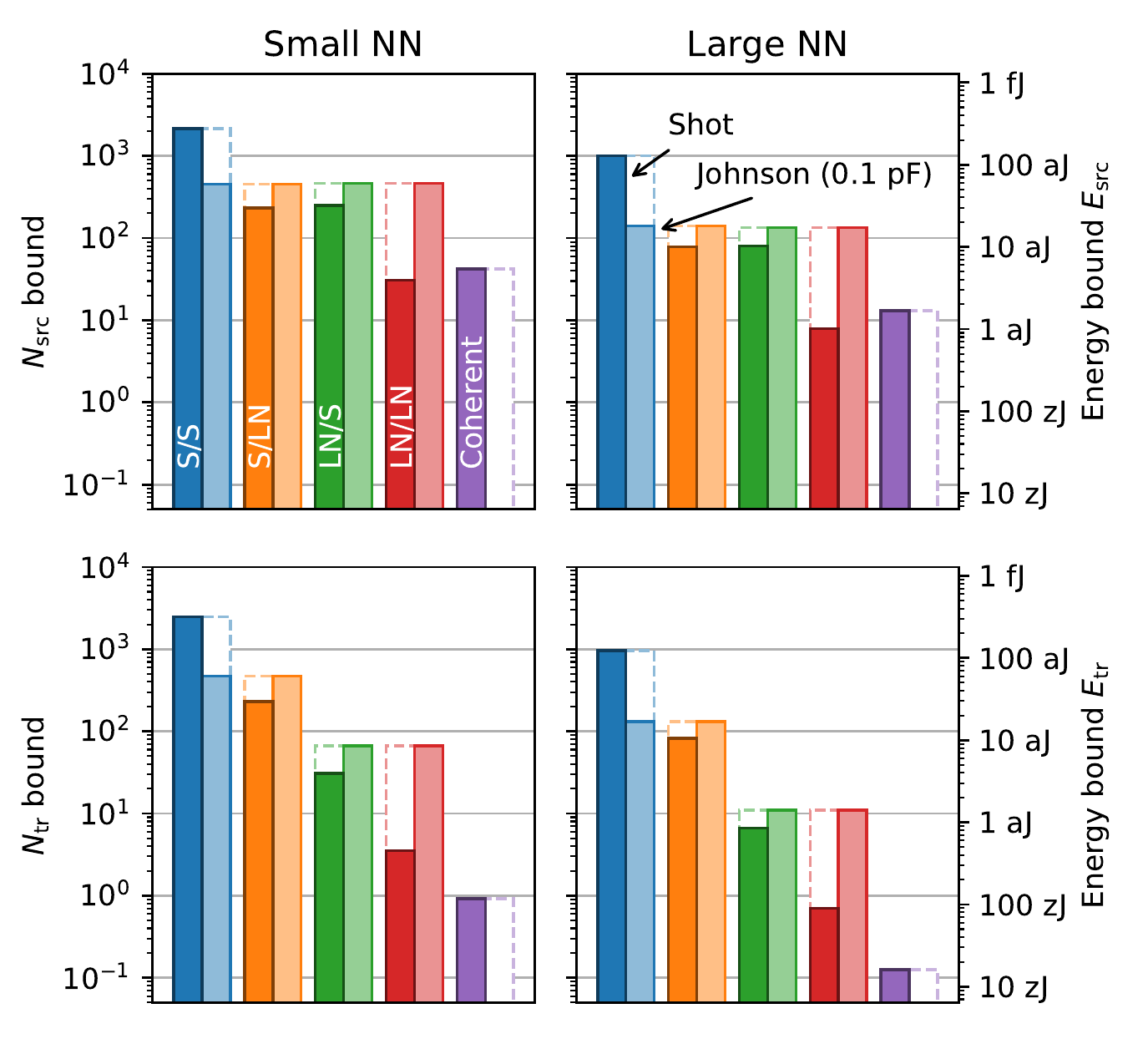}
\caption{Summary of lower limits to photon number per MAC (either $N_{\rm src}$ or $N_{\rm tr}$) due to shot noise (Fig.~\ref{fig:f3}) and Johnson noise (Fig.~\ref{fig:f4}).}
\label{fig:f5}
\end{center}
\end{figure}

Naturally, these noise terms will set a lower limit to the (optical) energy consumption of the Netcast receiver.  We can quantify this by running simulations of benchmark DNNs at a range of power levels to find the cutoff point.  As a benchmark, here we consider two three-layer fully-connected DNNs trained on the MNIST dataset: a ``small'' DNN of size $(28\!\times\!28)\rightarrow100\rightarrow100\rightarrow10$ and a ``large'' DNN of size $(28\!\times\!28)\rightarrow1000\rightarrow1000\rightarrow10$ (see Ref.~\cite{Hamerly2019} for details).  To simplify the analysis, the effects of Johnson and shot noise are studied separately.

accumulationFig.~\ref{fig:f3} considers shot noise and plots the DNN error rate as a function of photon number.  As discussed above, there are two ways to count photons: at the source $N_{\rm src}$ (left column) and at the transmitter output $N_{\rm tr}$ (right column).  We use the same $N_{\rm src}$ (resp.\ $N_{\rm tr})$ for each layer so that there is only one parameter to vary (rather than three), although layerwise optimization can yield better energy efficiency \cite{Garg2021}.  The expected behavior is observed: random guessing for $N \ll 1$, ``digital'' accuracy for $N \gg 1$, and a crossover that corresponds to the standard quantum limit (SQL) for Netcast running these DNNs.  The SQLs for the small and large DNNs differ by a factor of 2--5, similar to the trend observed for output-stationary photoelectric-multiplication \cite{Hamerly2019}.  As expected in Sec.~\ref{sec:var}, the quantum limit in the simple configuration (S/S) is quite poor ($\gtrsim 10^3$ photons/MAC), but the low-noise and coherent schemes have much better sensitivity, the SQL reduced up to $10^2\times$ relative to $N_{\rm src}$ and $10^4\times$ relative to $N_{\rm tr}$.  These reductions correlate with the factors $F_{\rm src}$, $F_{\rm tr}$ from Table~\ref{tab:t1}.

Fig.~\ref{fig:f4} performs a similar analysis with Johnson noise, which is only applicable to incoherent models.  Johnson noise scales with readout capacitance, which here we set to a conservative value of $C = 0.1$~pF.  Note that, per Eq.~(\ref{eq:sig}), Johnson noise depends only on $N_{\rm src}$; therefore, the curves of all designs overlap in the left column.  Relative to $N_{\rm tr}$, the the noise depends only on the transmitter, not the receiver.  Therefore, there are only two distinct cases when varying $N_{\rm tr}$.  We find a bound of $N_{\rm src} = 430$ (resp.\ 130) for the small (resp.\ large) DNN in the simple-transmitter case.  The low-noise transmitter can reduce the $N_{\rm tr}$ bound by about $10\times$, which makes sense because $\langle |w_{mn}| \rangle \approx 0.1$ for these neural networks.

From these figures, we now have lower bounds to the photon number (or equivalently optical energy) per MAC.  Putting this all together, Fig.~\ref{fig:f5} lists these bounds for each design (S/S, S/LN, LN/S, LN/LN, coherent), noise source (Johnson, shot), and accounting method (source power, transmitted power).  Several observations from this figure are worth emphasizing:

\begin{enumerate}
	\item The general ordering from least to most energy-efficient is: S/S $<$ S/LN $<$ LN/S $<$ LN/LN $<$ Coherent.
	\item The efficiency gains are most pronounced relative to $N_{\rm tr}$.
	\item S/LN and LN/S have similar performance relative to $N_{\rm src}$, but LN/S is much better relative to $N_{\rm tr}$.  Therefore, if we must economize and can only make one device low-noise, it should be the transmitter.
	\item Among the incoherent schemes, only S/S is shot-noise limited (at $C = 0.1$~pF).  S/LN and LN/S see roughly equal contributions from both noise sources, while LN/LN is strongly Johnson-noise limited.  This means that there is little reason to go from LN/S to LN/LN because both are bottlenecked by the (same) Johnson noise bound.  Reduced readout capacitance, detector avalanching, or input pre-amplification will be needed to reap the benefits of LN/LN's lower shot noise.
	\item The coherent scheme is consistently the best, even in the absence of Johnson noise.  The large NN here sets a new record: $\approx 15$~zJ/MAC.  This is lower than the 50--100~zJ/MAC predicted in Ref.~\cite{Hamerly2019} and the 250~zJ/MAC achieved in Ref.~\cite{Wang2022}.  High-fidelity computation at $< 1$ photon per MAC is possible because the computation result is the sum over many MACs, which can maintain an acceptable SNR even if the SNR of individual MACs is below unity \cite{Wang2022, Sludds2022}.
\end{enumerate}

\section{Throughput and Crosstalk}
\label{sec:tput}

If the client operates as a matrix-vector multiplier (as shown in Figs.~\ref{fig:f1}-\ref{fig:f2}), it will perform one MAC per weight received; thus the client's throughput is fundamentally limited by the link.  We can also envision matrix-matrix clients with on-chip fan-out after the PBS; this increases the maximum throughput by a constant factor (i.e.~$k$ MACs per weight) at the expense of complexity (receiver circuit is duplicated $k$ times over); nevertheless, link bandwidth still places a limit on throughput in this case.  Fundamentally, the channel capacity of an optical link is limited by crosstalk.  Since time and frequency are noncommuting operators, the time-frequency bins of Fig.~\ref{fig:f1}(a) are actually non-orthogonal, which will lead to inevitable crosstalk between the matrix elements.  This crosstalk ultimately limits the weight throughput of the channel.  

\subsection{Analytic Estimate}
\label{sec:est}

First, we create a simplified model in order to derive an estimate for the crosstalk.  For concreteness we focus on ring-based modulators and multiplexers (Fig.~\ref{fig:f2}(b-c)).  The crosstalk in this case takes two forms and can be estimated analytically:
\begin{enumerate}
	\item {\it Temporal crosstalk.}  This arises from the finite photon lifetime in the ring modulators and their finite $RC$ time constant.  Lumping these together, we define an approximate modulator response time $\tau = \sqrt{1/\kappa^2 + (RC)^2}$.  For efficient modulators, $RC \approx \kappa$, so $\tau \approx \sqrt{2}/\kappa$.  We take temporal crosstalk to have the form $\chi_{t} = e^{-T/\tau}$, where $T$ is the time between weights.  This sets an upper limit on the symbol rate $R = 1/T$ of the modulators:
	\beq
		R \leq \frac{\kappa}{\sqrt{2}\log(1/\chi_t)} = \frac{2\pi f_0}{\sqrt{2}\,Q \log(1/\chi_t)}
	\eeq
	where $f_0$ is the optical carrier frequency and $Q$ is the ring's $Q$-factor.
	
	\item {\it Frequency crosstalk.}  We will inevitably have crosstalk between channels of the WDM receiver (even if we had a ``perfect WDM'', the transmitter rings would have frequency crosstalk too).  This is set by the Lorentzian lineshape $\chi_\omega = (\kappa/2)^2/(\Delta \omega^2 + (\kappa/2)^2)$, where $\Delta\omega$ is the spacing between neighboring WDM channels.  In the low-crosstalk case $\Delta\omega \gg \kappa$, this gives a minimum channel spacing:
	\beq
		\Delta\omega \geq \frac{\kappa}{2\sqrt{\chi_\omega}} = \frac{2\pi f_0}{2Q \sqrt{\chi_\omega}}
	\eeq
\end{enumerate}
\begin{figure}[tbp]
\begin{center}
\includegraphics[width=1.00\columnwidth]{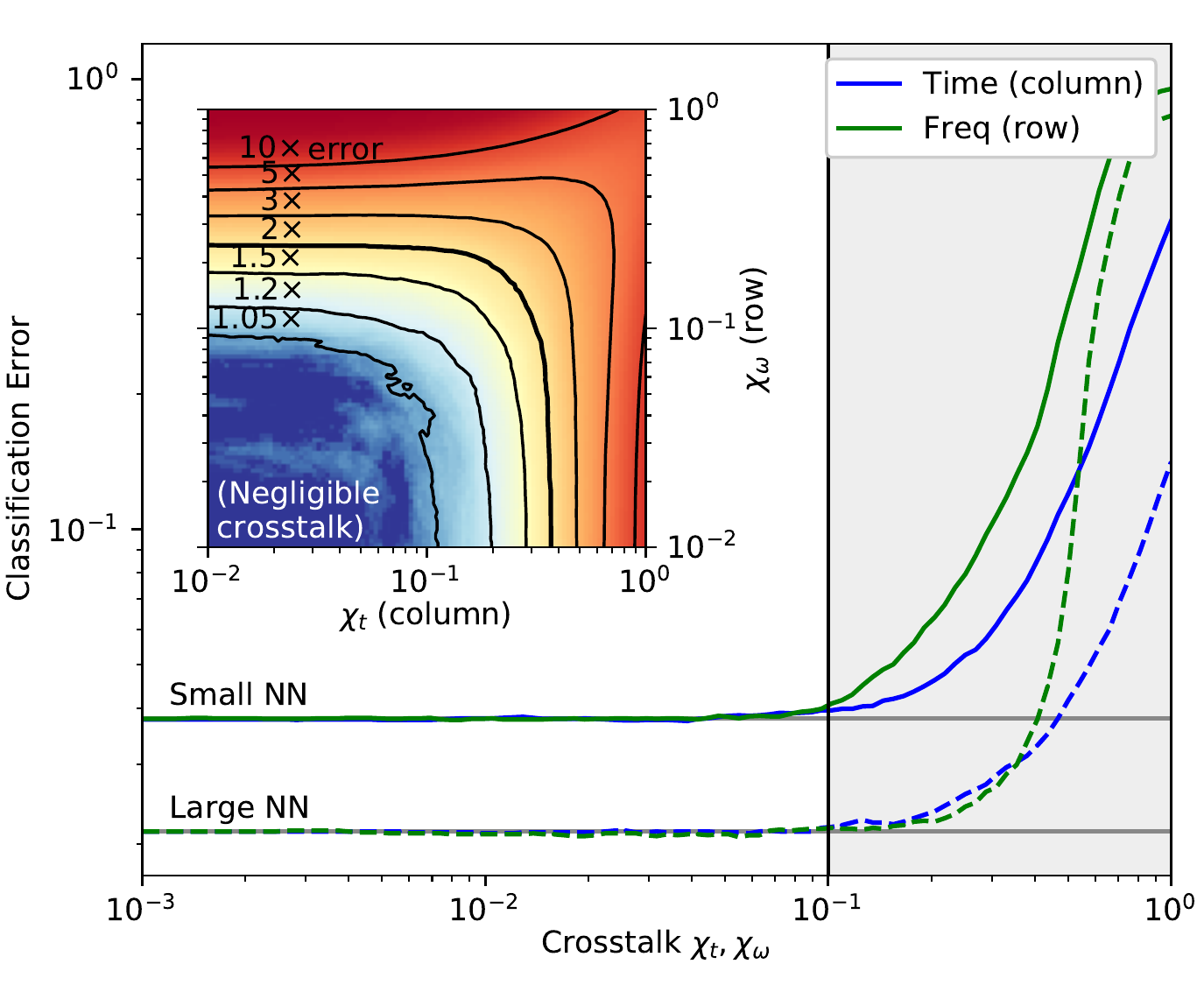}
\caption{Effect of crosstalk on MNIST DNN classification accuracy.  Crosstalk only degrades the neural network when $\chi_t, \chi_\omega \gtrsim 0.1$.  Inset: classification error in the presence of joint (time plus frequency) crosstalk for the small DNN.}
\label{fig:f6}
\end{center}
\end{figure}
To obtain a model for the effect of crosstalk, note that the detector charge $Q_m$ is a bilinear function of the weight signal $w(t) \leftrightarrow w_{mn}$ and the activation signal $x(t) \leftrightarrow x_n$.  The most general bilinear form is: $Q_m = \sum_{m',nn'} Y_{mm',nn'} w_{m'n'} x_n$.  The form of the tensor $Y$ is subject to a number of symmetries.  First, $Q_m$ is the charge accumulated after time integration, so time-translation symmetry should be respected: $Y_{mm',nn'} = Y_{mm',(n+p)(n'+p)}$.  Likewise, assuming the demultiplexer's frequency filters are the same for each channel, we also have frequency-translation symmetry: $Y_{mm',nn'} = Y_{(m+p)(m'+p),nn'}$.  Altogether, this means that the bilinear form reduces to convolution:
\beq
	Q_m = \sum_{pq,n} X_{pq} w_{m+p,n+q} x_n \label{eq:qmconv}
\eeq
$X_{pq}$ (normalized to $X_{00} = 1$) is the crosstalk matrix.  In useful cases, crosstalk is weak and the nearest-neighbor terms dominate: these are (1) the temporal (column) crosstalk $(X_{0,1}, X_{0,-1})$ and (2) the frequency (row) crosstalk $(X_{1,0}, X_{-1,0})$.  Often, the crosstalk is also symmetric ($X_{p,q} = X_{-p,-q}$), so there are only two independent parameters: $\chi_t = X_{0,1} = X_{0,-1}$ and $\chi_\omega = X_{1,0} = X_{-1,0}$.

Crosstalk must be sufficiently low for the DNN to function accurately.  Since Netcast relies on analog signals, tolerance may be more strict than in a comparable communications system.  Ref.~\cite{Hamerly2019IEDM} analyzed termporal crosstalk for simple MNIST DNNs \cite{Hamerly2019} and AlexNet \cite{Krizhevsky2012} and found that $\chi_t \lesssim 0.05$ is usually sufficient.  As Fig.~\ref{fig:f6} shows, spatial and joint crosstalk have a similar threshold; setting $\chi_t = \chi_\omega \equiv \chi$, the channel capacity will be bounded by:
\beq
	C = R \frac{2\pi B}{\Delta\omega} \leq \frac{2\pi\sqrt{2\chi}}{\log(1/\chi)} B \equiv C_0 B
\eeq
Here $B$ is the bandwidth (in Hz) and $C_0$ is the normalized symbol rate (units 1/Hz-s).  Table~\ref{tab:t2} shows the capacity as a function of crosstalk, both as a normalized rate and assuming use of the full C-band (1530--1565~nm, $B = 4.4$~THz), which can be converted to an equivalent digital data capacity, assuming 8-bits weights.  For reasonable crosstalk values, the data rates are comparable to those achieved with High Bandwidth Memory (HBM) links in workstation GPUs (6--12~Tbps \cite{Jia2018}).

\begin{table}[tbp]
\begin{center}
\caption{Estimate of maximum link bandwidth as a function of crosstalk.}
\begin{tabular}{c|ccc}
\hline\hline
Crosstalk $\chi$ & Symbol rate $C_0$ & Capacity $C$ (C-band) & $\times 8$~b/wt \\
\hline
0.1   & 1.22 & 5.3~Twt/s & 43~Tbps\\
0.05  & 0.66 & 2.9~Twt/s & 23~Tbps \\
0.01  & 0.19 & 850~Gwt/s & 6.8~Tbps \\
0.005 & 0.12 & 520~Gwt/s & 4.2~Tbps \\
0.001 & 0.04 & 180~Gwt/s & 1.2~Tbps \\ 
\hline\hline
\end{tabular}
\label{tab:t2}
\end{center}
\end{table}

\subsection{Full Model}
\label{sec:full}

A full model for crosstalk is always hardware-dependent.  In this section, we study a particular configuration---the coherent TIFS scheme where server-side modulation and client-side multiplexing are performed with microring resonators (Fig.~\ref{fig:f2}(c)).  To simplify the math, we assume that the modulators operate in the small-signal (linear) regime, as the addition of modulator nonlinearity is unlikely to affect the magnitude of the crosstalk.  Under these assumptions, the weight server output field $a_w(t)$ and the client LO output $a_x(t)$ are linear functions of $w_{mn}$ and $x_n$, respectively; as a result, the detector signal will be bilinear in $w_{mn}$ and $x_n$.  The symmetry assumptions discussed in Sec.~\ref{sec:est} lead to a convolutional crosstalk form Eq.~(\ref{eq:qmconv}), namely $Q_m = \sum_{pq,n} X_{pq} w_{m+p,n+q}x_n$.

%\begin{figure}[b!]
%\begin{center}
%\includegraphics[width=1.0\textwidth]{Note16-Fig6.pdf}
%\caption{Path of the optical field as it passes from the weight-server source to the client detector array.  The six steps are: (1) initialization at source, (2) pre-filtering by rings $m' < m$, (3) modulation by ring $m$, (4) output filtering by ring $m$, (5) post-filtering by rings $m' > m$, and (6) wavelength demultiplexing.}
%\label{fig:f6}
%\end{center}
%\end{figure}

It remains to compute the crosstalk kernel $X_{pq}$.  As before, we will focus here on dominant nearest-neighbor terms.  We find $X_{pq}$ by following the field as it passes from source to detector, in the presence of a {\it single} data pulse $w_{mn}$ (the overall field is a linear combination of such contributions).  This is a sequence of six steps, illustrated in Fig.~\ref{fig:f6}:

\begin{figure}[htbp]
\begin{center}
\includegraphics[width=1.00\columnwidth]{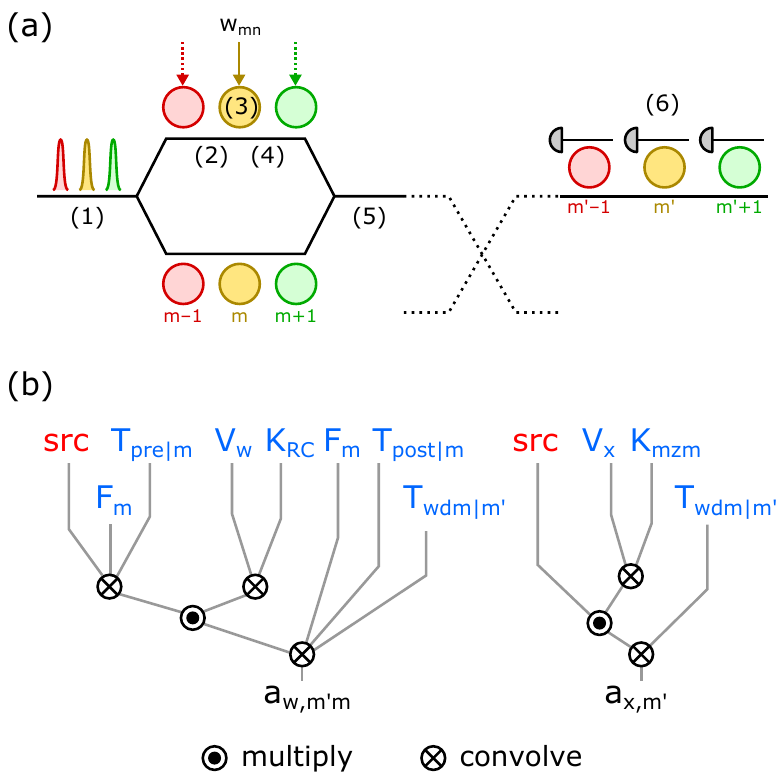}
\caption{(a) Path of the optical field as it passes from the weight-server source to the client detector array.  The six steps are: initialization at source, pre-filtering by rings $m' < m$, modulation by ring $m$, output filtering by ring $m$, post-filtering by rings $m' > m$, and wavelength demultiplexing.  (b) Computation of detector-side fields $a_{w}(\omega|m'm)$ and $a_{x}(\omega|m'm)$.}
\label{fig:f7}
\end{center}
\end{figure}

\begin{enumerate}
	\item {\it Initialization.}  We start with a frequency comb so that the input field consists of a discrete sum of comb lines:
	$a^{(1)}(\omega) = \sum_{m'} \delta(\omega - m'\Omega)$
\item {\it Pre-filtering (rings $m' < m$).}  The field is divided into two paths in the WDM-MZM.  These paths impart opposite perturbations, so it is sufficient to consider only the top path.  The light passes through an array of unperturbed rings.  Assume that these rings are all critically coupled (which makes sense from a robustness point of view because it makes the WDM-MZM less sensitive to imperfect splitting ratios).  The transfer function for a critically coupled ring is given by $T(\Delta\omega) = -i\,\Delta\omega/(\kappa - i\,\Delta\omega)$.  As a result, the field at this stage is:
	\begin{align}
		a^{(2)}(\omega|m) & = \underbrace{\Bigl(\prod_{m'<m} T(\omega - m'\Omega)\Bigr)}_{T_{\rm pre}(\omega | m)} a^{(1)}(\omega) \nonumber \\
		& = \sum_{m'} T_{\rm pre}(m'\Omega | m) \delta(\omega - m'\Omega)
	\end{align}
	\item {\it Modulator (ring $m$).}  The modulator's internal field is obtained by solving the equations:
	\bea
		\dot{a}(t) & = & \bigl(-\kappa - i\Delta(t)\bigr) a(t) - \sqrt{\kappa}\,a_{\rm in}(t) \nonumber \\
		a_{\rm out}(t) & = & \sqrt{\kappa}\,a(t) + a_{\rm in}(t)
	\eea
	Here $\Delta(t)$, the detuning induced by the signal, is treated in the linear regime as a perturbation.  The field in the ring can be expanded to a constant and perturbed term: $a^{(3)}(t) + \delta a^{(3)}(t)$ (or $a^{(3)}(\omega) + \delta a^{(3)}(\omega)$).  The unperturbed part is:
	\beq
		a^{(3)}(\omega|m) = -\frac{1}{\sqrt{\kappa}}\underbrace{\frac{\kappa}{\kappa - i(\omega-m\Omega)}}_{F(\omega - m\Omega)} a^{(2)}(\omega|m) \label{eq:a3}
	\eeq
	\item {\it Modulator Perturbation.}  The equation for the perturbed part is:
	\beq
		\delta\dot{a}(t) = (-\kappa-im\Omega)\delta a(t) \underbrace{-\, i \Delta(t) a(t)}_{+\kappa S(t)}  \label{eq:dapert}
	\eeq
	where we have defined the source term
	\bea
		S(t|m) & = & \kappa^{-1} \Delta(t) a^{(3)}(t|m) \nonumber \\
		S(\omega|m) & = & \kappa^{-1} [\Delta(\omega) \otimes a^{(3)}(\omega|m)]
	\eea
	and the detuning $\Delta$ is the modulator input waveform $V_w(t)$ passed through its RC filter $K_{RC}(t)$:
	\beq
		\Delta(t) = [V_w \otimes K_{RC}](t)
		\ \ \Leftrightarrow\ \ 
		\Delta(\omega) = V_w(\omega) K_{RC}(\omega)
	\eeq
	With $S(\omega|m)$ in hand, we solve Eq.~(\ref{eq:dapert}) to find $\delta a^{(3)}(\omega|m) = F(\omega - m\Omega) S(\omega|m)$ with the kernel $F(\omega)$ defined in Eq.~(\ref{eq:a3}).
	
	We only care about the perturbation term from here on, since the unperturbed field cancels out due to the MZM's dual drive.  Since there is no perturbation term in the modulator input, the perturbation to the modulator output $\delta a^{(4)}$ is just:
	\beq
		\delta a^{(4)}(\omega|m) = \sqrt{\kappa}\,\delta a^{(3)}(\omega|m) = \sqrt{\kappa}\,F(\omega - m\Omega) S(\omega|m)
	\eeq
	\item {\it Post-filtering (rings $m' > m$).}  The perturbed field passes through the rest of the rings and is accordingly filtered.  As mentioned earlier, the unperturbed term is cancelled by the symmetry of the WDM-MZM, so we can write $a^{(5)}$ instead of $\delta a^{(5)}$ from here on.
	\beq
		a^{(5)}(\omega|m) = \underbrace{\Bigl(\prod_{m'>m} T(\omega - m'\Omega)\Bigr)}_{T_{\rm post}(\omega | m)} \delta a^{(4)}(\omega|m)
	\eeq
	\item {\it Demultiplexing.}  Let's consider detector $m'$.  After mixing with the local oscillator, the field passes by all rings with $m'' < m'$ and is reflected to ring the through port of ring $m'$, where it goes into a detector.  The amplitude in this detector (denoted $a_w$ because of it originates from the weight signal) will be:
	\begin{align}
		& a_w^{(6)}(\omega|m'm) \nonumber \\
		& = \biggl[\underbrace{F(\omega - m'\Omega) \prod_{m''<m'} T(\omega - m''\Omega)}_{T_{\rm wdm}(\omega|m')}\biggr] a^{(5)}(\omega|m)
	\end{align}
	Note that, for an infinite array of equally spaced frequencies, by symmetry $a_w^{(6)}(\omega|m'm)$ depends only on the difference index $(m'-m)$.
\end{enumerate}
Similarly, we can work through the local oscillator.  Here a broadband MZM is used.  The input waveform $V_x(t)$ is convolved by the MZM's filter $K_{\rm mzm}(t)$ to produce the modulation amplitude $\theta(t)$ (here assuming $\theta \ll 1$).  This is multiplied by the input frequency comb (same as $a^{(1)}(\omega)$) and sent through the same client-side WDM filter.  The final result (in frequency space) is:
\begin{align}
	& a^{(6)}_x(\omega|m') \nonumber \\
	& = T_{\rm wdm}(\omega|m')\Bigl(\bigl(V_x(\omega) K_{\rm mzm}(\omega)\bigr) \otimes \sum_{m} \delta(\omega - m\Omega)\Bigr)
\end{align}
Given the form Eq.~(\ref{eq:qmconv}), the crosstalk element $X_{pq}$ is the interference at detector $m$ between the signal from $x_n$ (pulse at the current time step) and the signal from $w_{m+p,n+q}$ (pulse $q$ time steps in the future, from modulator $m+p$).  This is:
\beq
	X_{pq} \propto \mbox{Re}\bigl[a_x^{(6)}(\omega|m)^* a_w^{(6)}(\omega|m,m+p) e^{iq\omega T} \bigr]
\eeq

%\begin{figure}[t!]
%\begin{center}
%\includegraphics[width=0.8\textwidth]{Note16-Fig7.pdf}
%\caption{Time (row) and frequency (column) crosstalk as a function of time and frequency spacing.  Since crosstalk is asymmetric, the figure plots averages ($\tfrac{1}{2}(X_{+1,0} + X_{-1,0})$ for frequency crosstalk and $\tfrac{1}{2}(X_{0,+1} + X_{0,-1})$ for temporal crosstalk).}
%\label{fig:f7}
%\end{center}
%\end{figure}
%
%\begin{figure}[t!]
%\begin{center}
%\includegraphics[width=1.0\textwidth]{Note16-Fig8.pdf}
%\caption{Left: MNIST classification error on the small NN as a function of time and frequency spacing.  Right: corresponding plot of error as a function of normalized bandwidth, showing the capacity limit $C_{\rm max} \approx 0.25$.}
%\label{fig:f8}
%\end{center}
%\end{figure}

\begin{figure}[t!]
\begin{center}
\includegraphics[width=1.00\columnwidth]{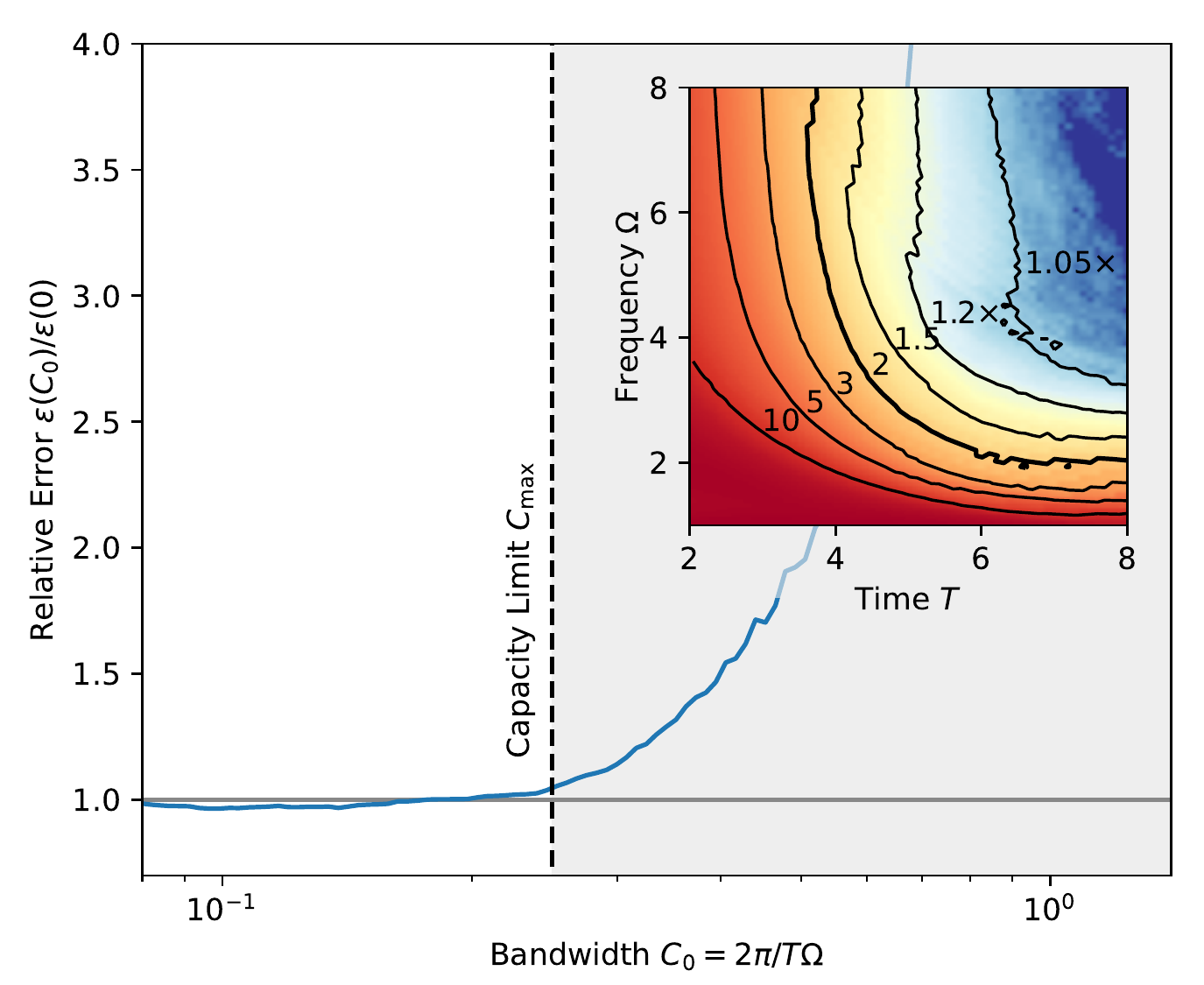}
\caption{Bandwidth-accuracy tradeoff for the microring-based Netcast implementation studied in Sec.~\ref{sec:full}.  The classification error shows a noticeable jump beyond the capacity limit $C_{\rm max} \approx 0.25$.  Inset: relative error as a function of time- and frequency-spacing $(T, \Omega)$, where the capacity is given by $C_0 = 2\pi/T\Omega$.}
\label{fig:f8}
\end{center}
\end{figure}

These matrix elements depend on many variables including the pulse shapes $V_{x,w}(t)$, the RC time constant, the modulator and WDM photon lifetimes, etc.  For simplicity, here we consider the case of (1) square-wave $V_{x,w}$ with duty cycle $\tfrac{1}{2}$, (2) $RC = \kappa$ for modulator rings, and (3) identical $\kappa$ for modulator and WDM rings.  These parameters set, the matrix elements $X_{pq}$ depend only on the time- and frequency-spacing $(T, \Omega)$.  Given the crosstalk matrix, we compute the MNIST classification accuracy (small NN) as a function of $(T, \Omega)$, from which one can derive the optimal accuracy as a function of the capacity $C_0 = 2\pi/T\Omega$.  This is plotted in Fig.~\ref{fig:f8}.  The observed capacity limit $C_{\rm max} = 0.25$ is within a factor of $2.5\times$ of our analytic estimate obtained in Table~\ref{tab:t2}.  It should not be too surprising that this is somewhat smaller than the analytic value, as the model used here contains a larger number of bandwidth-limiting factors that can induce additional crosstalk; moreover, the variable choices above (square waves, $RC=\kappa$, etc.) could likely be further optimized.

\section{Conclusion}

As computing moves to the edge, optics can open up new possibilities to deliver high performance while simultaneously adhering to strict SWaP constraints.  In recent work, we have introduced \cite{Netcast1} and experimentally demonstrated \cite{Sludds2022} NetCast, a photonic server-client architecture that leverages unique advantages of optics---the high bandwidth of optical links, support for wavelength division multiplexing, wavelength-parallel modulation, and analog integration detection---to split the DNN inference problem into two complementary tasks: weight encoding at the server and lightweight optical postprocessing at the client.  This approach effectively pushes the energy- and memory-intensive tasks to the server, significantly relieving pressure on the client's SWaP constraints.

This paper has analyzed the limits to two factors the govern the performance of Netcast: energy efficiency and throughput.  By pushing the weight retrieval problem to the server, this protocol allows the electrical energy consumption for a MVM to be reduced from $O(N^2)$ to in principle $O(N)$.  Another important consideration is the optical energy, particularly for situations employing large-scale fan-out to many clients or deployment over long-distance links.  Optical energy consumption is closely tied to the need to maintain a sufficient SNR in the presence of Johnson and shot noise.  We analyzed five unique server-client configurations which offer tradeoffs between hardware complexity and sensitivity to low optical powers.  Numerical simulations reveal that the simplest (differential signaling) design requires $>10^3$~photons/MAC for accurate inference, while the most complex (coherent detection) can function with as low as 0.1~photons/MAC.

Throughput of the Netcast client is limited by the delivery of optical weights.  Since the weights are encoded in the analog domain, time- and frequency-domain crosstalk pose a fundamental limit to throughput.  We have derived analytic expressions for these limits that show Netcast should data bandwidths comparable to high-end GPU HBM, provided the full optical C-band is used.  A more detailed physical model agrees roughly with these estimates.  Improvements in throughput are possible using coherent detection (a $4\times$ factor due to use of quadrature and polarization diversity), operation beyond the C-band (which may be enabled by advances in frequency combs \cite{Geng2022} and novel amplifiers \cite{Maes2022}), and spatial multiplexing.

The high theoretical performance limits of Netcast support the use of optics in edge-computing situations where the server and client are connected by an optical link.  More broadly, they highlight an exciting new possibility for computing and communications: the use of broadband {\it analog} optical interconnects to accelerate distributed computing tasks.  These interconnects may supplement existing digital links, harnessing the innate parallelism of analog optics to enable new computational architectures in DNN inference, training, sensing, data fusion, and distributed intelligence.

\section*{Acknowledgment}

This research was funded by NTT Research Inc.\ and NSF EAGER (CNS-1946976).  A.S.\ and S.B.\ are supported by NSF Graduate Research Fellowships.  L.B.\ is supported by an NSERC Postgraduate Fellowship.

\ifCLASSOPTIONcaptionsoff
  \newpage
\fi



\bibliographystyle{IEEEtran}
\bibliography{IEEEabrv,refs.bib}

\begin{thebibliography}{10}
\providecommand{\url}[1]{#1}
\csname url@samestyle\endcsname
\providecommand{\newblock}{\relax}
\providecommand{\bibinfo}[2]{#2}
\providecommand{\BIBentrySTDinterwordspacing}{\spaceskip=0pt\relax}
\providecommand{\BIBentryALTinterwordstretchfactor}{4}
\providecommand{\BIBentryALTinterwordspacing}{\spaceskip=\fontdimen2\font plus
\BIBentryALTinterwordstretchfactor\fontdimen3\font minus
  \fontdimen4\font\relax}
\providecommand{\BIBforeignlanguage}[2]{{%
\expandafter\ifx\csname l@#1\endcsname\relax
\typeout{** WARNING: IEEEtran.bst: No hyphenation pattern has been}%
\typeout{** loaded for the language `#1'. Using the pattern for}%
\typeout{** the default language instead.}%
\else
\language=\csname l@#1\endcsname
\fi
#2}}
\providecommand{\BIBdecl}{\relax}
\BIBdecl

\bibitem{Yu2017}
W.~Yu, F.~Liang, X.~He, W.~G. Hatcher, C.~Lu, J.~Lin, and X.~Yang, ``A survey
  on the edge computing for the internet of things,'' \emph{IEEE Access},
  vol.~6, pp. 6900--6919, 2017.

\bibitem{Sze2017}
V.~Sze, Y.-H. Chen, T.-J. Yang, and J.~S. Emer, ``Efficient processing of deep
  neural networks: A tutorial and survey,'' \emph{Proceedings of the IEEE},
  vol. 105, no.~12, pp. 2295--2329, 2017.

\bibitem{Chen2014}
T.~Chen, Z.~Du, N.~Sun, J.~Wang, C.~Wu, Y.~Chen, and O.~Temam, ``{DianNao}: A
  small-footprint high-throughput accelerator for ubiquitous
  machine-learning,'' \emph{ACM Sigplan Notices}, vol.~49, no.~4, pp. 269--284,
  2014.

\bibitem{Howard2017}
A.~G. Howard, M.~Zhu, B.~Chen, D.~Kalenichenko, W.~Wang, T.~Weyand,
  M.~Andreetto, and H.~Adam, ``Mobilenets: Efficient convolutional neural
  networks for mobile vision applications,'' \emph{arXiv preprint
  arXiv:1704.04861}, 2017.

\bibitem{Iandola2016}
F.~N. Iandola, S.~Han, M.~W. Moskewicz, K.~Ashraf, W.~J. Dally, and K.~Keutzer,
  ``{SqueezeNet}: {AlexNet}-level accuracy with 50x fewer parameters and< 0.5
  mb model size,'' \emph{arXiv preprint arXiv:1602.07360}, 2016.

\bibitem{Xu2018}
X.~Xu, Y.~Ding, S.~X. Hu, M.~Niemier, J.~Cong, Y.~Hu, and Y.~Shi, ``Scaling for
  edge inference of deep neural networks,'' \emph{Nature Electronics}, vol.~1,
  no.~4, pp. 216--222, 2018.

\bibitem{Szegedy2016}
C.~Szegedy, V.~Vanhoucke, S.~Ioffe, J.~Shlens, and Z.~Wojna, ``Rethinking the
  inception architecture for computer vision,'' in \emph{Proceedings of the
  IEEE conference on computer vision and pattern recognition}, 2016, pp.
  2818--2826.

\bibitem{Dai2019}
Z.~Dai, Z.~Yang, Y.~Yang, J.~Carbonell, Q.~V. Le, and R.~Salakhutdinov,
  ``{Transformer-XL}: Attentive language models beyond a fixed-length
  context,'' \emph{arXiv preprint arXiv:1901.02860}, 2019.

\bibitem{Krestinskaya2019}
O.~Krestinskaya, A.~P. James, and L.~O. Chua, ``Neuromemristive circuits for
  edge computing: A review,'' \emph{IEEE transactions on neural networks and
  learning systems}, vol.~31, no.~1, pp. 4--23, 2019.

\bibitem{Ambrogio2018}
S.~Ambrogio, P.~Narayanan, H.~Tsai, R.~M. Shelby, I.~Boybat, C.~Di~Nolfo,
  S.~Sidler, M.~Giordano, M.~Bodini, N.~C. Farinha \emph{et~al.},
  ``Equivalent-accuracy accelerated neural-network training using analogue
  memory,'' \emph{Nature}, vol. 558, no. 7708, pp. 60--67, 2018.

\bibitem{Paek1987}
E.~G. Paek and D.~Psaltis, ``Optical associative memory using {Fourier}
  transform holograms,'' \emph{Optical Engineering}, vol.~26, no.~5, p. 265428,
  1987.

\bibitem{New2017}
N.~J. New, ``Reconfigurable optical processing system,'' Mar.~14 2017, {US}
  {P}atent 9,594,394.

\bibitem{Shen2017}
Y.~Shen, N.~C. Harris, S.~Skirlo, M.~Prabhu, T.~Baehr-Jones, M.~Hochberg,
  X.~Sun, S.~Zhao, H.~Larochelle, D.~Englund \emph{et~al.}, ``Deep learning
  with coherent nanophotonic circuits,'' \emph{Nature Photonics}, vol.~11,
  no.~7, p. 441, 2017.

\bibitem{Prabhu2020}
M.~Prabhu, C.~Roques-Carmes, Y.~Shen, N.~Harris, L.~Jing, J.~Carolan,
  R.~Hamerly, T.~Baehr-Jones, M.~Hochberg, V.~{\v{C}}eperi{\'c} \emph{et~al.},
  ``Accelerating recurrent ising machines in photonic integrated circuits,''
  \emph{Optica}, vol.~7, no.~5, pp. 551--558, 2020.

\bibitem{Tait2017}
A.~N. Tait, T.~F. Lima, E.~Zhou, A.~X. Wu, M.~A. Nahmias, B.~J. Shastri, and
  P.~R. Prucnal, ``Neuromorphic photonic networks using silicon photonic weight
  banks,'' \emph{Scientific Reports}, vol.~7, no.~1, p. 7430, 2017.

\bibitem{Pai2022}
S.~Pai, Z.~Sun, T.~W. Hughes, T.~Park, B.~Bartlett, I.~A. Williamson,
  M.~Minkov, M.~Milanizadeh, N.~Abebe, F.~Morichetti \emph{et~al.},
  ``Experimentally realized in situ backpropagation for deep learning in
  nanophotonic neural networks,'' \emph{arXiv preprint arXiv:2205.08501}, 2022.

\bibitem{Bernstein2022}
L.~Bernstein, A.~Sludds, C.~Panuski, S.~Trajtenberg-Mills, R.~Hamerly, and
  D.~Englund, ``Single-shot optical neural network,'' \emph{arXiv preprint
  arXiv:2205.09103}, 2022.

\bibitem{Fang2019}
M.~Y.-S. Fang, S.~Manipatruni, C.~Wierzynski, A.~Khosrowshahi, and M.~R.
  DeWeese, ``Design of optical neural networks with component imprecisions,''
  \emph{Optics Express}, vol.~27, no.~10, pp. 14\,009--14\,029, 2019.

\bibitem{SaumilPaper}
S.~Bandyopadhyay, R.~Hamerly, and D.~Englund, ``Hardware error correction for
  programmable photonics,'' \emph{arXiv preprint arXiv:2103.04993}, 2021.

\bibitem{RyanPaper1}
R.~Hamerly, S.~Bandyopadhyay, and D.~Englund, ``Stability of self-configuring
  large multiport interferometers,'' \emph{arXiv preprint arXiv:2106.04363},
  2021.

\bibitem{RyanPaper2}
------, ``Accurate self-configuration of rectangular multiport
  interferometers,'' \emph{arXiv preprint arXiv:2106.03249}, 2021.

\bibitem{RyanPaper3}
------, ``Infinitely scalable multiport interferometers,'' \emph{arXiv preprint
  arXiv:2109.05367}, 2021.

\bibitem{Alexiev2021}
C.~Alexiev, J.~C. Mak, W.~D. Sacher, and J.~K. Poon, ``Calibrating rectangular
  interferometer meshes with external photodetectors,'' \emph{OSA Continuum},
  vol.~4, no.~11, pp. 2892--2904, 2021.

\bibitem{Netcast1}
R.~Hamerly, A.~Sludds, S.~Bandyopadhyay, L.~Bernstein, Z.~Chen, M.~Ghobadi, and
  D.~Englund, ``Edge computing with optical neural networks via {WDM} weight
  broadcasting,'' in \emph{Emerging Topics in Artificial Intelligence (ETAI)
  2021}, vol. 11804.\hskip 1em plus 0.5em minus 0.4em\relax SPIE, 2021, pp.
  55--60.

\bibitem{Sludds2022}
A.~Sludds, S.~Bandyopadhyay, Z.~Chen, Z.~Zhong, J.~Cochrane, L.~Bernstein,
  D.~Bunandar, P.~B. Dixon, S.~Hamilton, M.~Streshinsky \emph{et~al.},
  ``Delocalized photonic deep learning on the internet's edge,'' \emph{arXiv
  preprint arXiv:2203.05466}, 2022.

\bibitem{Zhong2021}
Z.~Zhong, W.~Wang, M.~Ghobadi, A.~Sludds, R.~Hamerly, L.~Bernstein, and
  D.~Englund, ``Ioi: In-network optical inference,'' in \emph{Proceedings of
  the ACM SIGCOMM 2021 Workshop on Optical Systems}, 2021, pp. 18--22.

\bibitem{Timurdogan2014}
E.~Timurdogan, C.~M. Sorace-Agaskar, J.~Sun, E.~S. Hosseini, A.~Biberman, and
  M.~R. Watts, ``An ultralow power athermal silicon modulator,'' \emph{Nature
  Communications}, vol.~5, p. 4008, 2014.

\bibitem{Stojanovic2018}
V.~Stojanovi{\'c}, R.~J. Ram, M.~Popovi{\'c}, S.~Lin, S.~Moazeni, M.~Wade,
  C.~Sun, L.~Alloatti, A.~Atabaki, F.~Pavanello \emph{et~al.}, ``Monolithic
  silicon-photonic platforms in state-of-the-art cmos soi processes,''
  \emph{Optics Express}, vol.~26, no.~10, pp. 13\,106--13\,121, 2018.

\bibitem{Haffner2018}
C.~Haffner, D.~Chelladurai, Y.~Fedoryshyn, A.~Josten, B.~Baeuerle, W.~Heni,
  T.~Watanabe, T.~Cui, B.~Cheng, S.~Saha \emph{et~al.}, ``Low-loss
  plasmon-assisted electro-optic modulator,'' \emph{Nature}, vol. 556, no.
  7702, p. 483, 2018.

\bibitem{DeCea2021}
M.~de~Cea, A.~Atabaki, and R.~Ram, ``Energy harvesting optical modulators with
  sub-attojoule per bit electrical energy consumption,'' \emph{Nature
  communications}, vol.~12, no.~1, pp. 1--9, 2021.

\bibitem{Rizzo2021}
A.~Rizzo, A.~Novick, V.~Gopal, B.~Y. Kim, X.~Ji, S.~Daudlin, Y.~Okawachi,
  Q.~Cheng, M.~Lipson, A.~L. Gaeta \emph{et~al.}, ``Integrated kerr frequency
  comb-driven silicon photonic transmitter,'' \emph{arXiv preprint
  arXiv:2109.10297}, 2021.

\bibitem{Zhang2022}
W.~Zhang, C.~Huang, H.-T. Peng, S.~Bilodeau, A.~Jha, E.~Blow, T.~F. de~Lima,
  B.~J. Shastri, and P.~Prucnal, ``Silicon microring synapses enable photonic
  deep learning beyond 9-bit precision,'' \emph{Optica}, vol.~9, no.~5, pp.
  579--584, 2022.

\bibitem{Hamerly2019}
R.~Hamerly, L.~Bernstein, A.~Sludds, M.~Solja{\v{c}}i{\'c}, and D.~Englund,
  ``Large-scale optical neural networks based on photoelectric
  multiplication,'' \emph{Physical Review X}, vol.~9, no.~2, p. 021032, 2019.

\bibitem{Spectrum}
R.~Hamerly, ``The future of deep learning is photonic: reducing the energy
  needs of neural networks might require computing with light,'' \emph{IEEE
  Spectrum}, vol.~58, no.~7, pp. 30--47, 2021.

\bibitem{Geng2022}
Y.~Geng, H.~Zhou, X.~Han, W.~Cui, Q.~Zhang, B.~Liu, G.~Deng, Q.~Zhou, and
  K.~Qiu, ``Coherent optical communications using coherence-cloned kerr soliton
  microcombs,'' \emph{Nature communications}, vol.~13, no.~1, pp. 1--8, 2022.

\bibitem{Kemal2016}
J.~N. Kemal, J.~Pfeifle, P.~Marin-Palomo, M.~D.~G. Pascual, S.~Wolf, F.~Smyth,
  W.~Freude, and C.~Koos, ``Multi-wavelength coherent transmission using an
  optical frequency comb as a local oscillator,'' \emph{Optics express},
  vol.~24, no.~22, pp. 25\,432--25\,445, 2016.

\bibitem{Jang2018}
J.~K. Jang, A.~Klenner, X.~Ji, Y.~Okawachi, M.~Lipson, and A.~L. Gaeta,
  ``Synchronization of coupled optical microresonators,'' \emph{Nature
  Photonics}, vol.~12, no.~11, pp. 688--693, 2018.

\bibitem{Liao2019}
P.~Liao, C.~Bao, A.~Almaiman, A.~Kordts, M.~Karpov, M.~H.~P. Pfeiffer,
  L.~Zhang, F.~Alishahi, Y.~Cao, K.~Zou \emph{et~al.}, ``Demonstration of
  multiple kerr-frequency-comb generation using different lines from another
  kerr comb located up to 50 km away,'' \emph{Journal of Lightwave Technology},
  vol.~37, no.~2, pp. 579--584, 2019.

\bibitem{Miller2012}
D.~A. Miller, ``Energy consumption in optical modulators for interconnects,''
  \emph{Optics Express}, vol.~20, no. 102, pp. A293--A308, 2012.

\bibitem{Atabaki2018}
A.~H. Atabaki, S.~Moazeni, F.~Pavanello, H.~Gevorgyan, J.~Notaros, L.~Alloatti,
  M.~T. Wade, C.~Sun, S.~A. Kruger, H.~Meng \emph{et~al.}, ``Integrating
  photonics with silicon nanoelectronics for the next generation of systems on
  a chip,'' \emph{Nature}, vol. 556, no. 7701, p. 349, 2018.

\bibitem{Jonsson2011}
B.~E. Jonsson, ``An empirical approach to finding energy efficient {ADC}
  architectures,'' in \emph{Proc. of 2011 IMEKO IWADC \& IEEE ADC Forum}, 2011,
  pp. 1--6.

\bibitem{Cosemans2019}
S.~Cosemans, B.~Verhoef, J.~Doevenspeck, I.~Papistas, F.~Catthoor, P.~Debacker,
  A.~Mallik, and D.~Verkest, ``Towards {10000TOPS/W} {DNN} inference with
  analog in-memory computing--a circuit blueprint, device options and
  requirements,'' in \emph{2019 IEEE International Electron Devices Meeting
  (IEDM)}.\hskip 1em plus 0.5em minus 0.4em\relax IEEE, 2019, pp. 22--2.

\bibitem{Tait2022}
A.~N. Tait, ``Quantifying power in silicon photonic neural networks,''
  \emph{Physical Review Applied}, vol.~17, no.~5, p. 054029, 2022.

\bibitem{Cole2021}
C.~Cole, ``Optical and electrical programmable computing energy use
  comparison,'' \emph{Optics Express}, vol.~29, no.~9, pp. 13\,153--13\,170,
  2021.

\bibitem{Horowitz2014}
M.~Horowitz, ``Computing's energy problem (and what we can do about it),'' in
  \emph{Solid-State Circuits Conference Digest of Technical Papers (ISSCC),
  2014 IEEE International}.\hskip 1em plus 0.5em minus 0.4em\relax IEEE, 2014,
  pp. 10--14.

\bibitem{Jouppi2017}
N.~P. Jouppi, C.~Young, N.~Patil, D.~Patterson, G.~Agrawal, R.~Bajwa, S.~Bates,
  S.~Bhatia, N.~Boden, A.~Borchers \emph{et~al.}, ``In-datacenter performance
  analysis of a tensor processing unit,'' in \emph{Computer Architecture
  (ISCA), 2017 ACM/IEEE 44th Annual International Symposium on}.\hskip 1em plus
  0.5em minus 0.4em\relax IEEE, 2017, pp. 1--12.

\bibitem{Reuther2019}
A.~Reuther, P.~Michaleas, M.~Jones, V.~Gadepally, S.~Samsi, and J.~Kepner,
  ``Survey and benchmarking of machine learning accelerators,'' in \emph{2019
  IEEE high performance extreme computing conference (HPEC)}.\hskip 1em plus
  0.5em minus 0.4em\relax IEEE, 2019, pp. 1--9.

\bibitem{Garg2021}
S.~Garg, J.~Lou, A.~Jain, and M.~Nahmias, ``Dynamic precision analog computing
  for neural networks,'' \emph{arXiv preprint arXiv:2102.06365}, 2021.

\bibitem{Wang2022}
T.~Wang, S.-Y. Ma, L.~G. Wright, T.~Onodera, B.~C. Richard, and P.~L. McMahon,
  ``An optical neural network using less than 1 photon per multiplication,''
  \emph{Nature Communications}, vol.~13, no.~1, pp. 1--8, 2022.

\bibitem{Hamerly2019IEDM}
R.~Hamerly, A.~Sludds, L.~Bernstein, M.~Prabhu, C.~Roques-Carmes, J.~Carolan,
  Y.~Yamamoto, M.~Solja{\v{c}}i{\'c}, and D.~Englund, ``Towards large-scale
  photonic neural-network accelerators,'' in \emph{2019 IEEE international
  electron devices meeting (IEDM)}.\hskip 1em plus 0.5em minus 0.4em\relax
  IEEE, 2019, pp. 22--8.

\bibitem{Krizhevsky2012}
A.~Krizhevsky, I.~Sutskever, and G.~E. Hinton, ``Imagenet classification with
  deep convolutional neural networks,'' in \emph{Advances in neural information
  processing systems}, 2012, pp. 1097--1105.

\bibitem{Jia2018}
Z.~Jia, M.~Maggioni, B.~Staiger, and D.~P. Scarpazza, ``Dissecting the {NVIDIA}
  {Volta} {GPU} architecture via microbenchmarking,'' \emph{arXiv preprint
  arXiv:1804.06826}, 2018.

\bibitem{Maes2022}
F.~Maes, M.~Sharma, L.~Wang, and Z.~Jiang, ``High power bdf/edf hybrid
  amplifier providing 27 db gain over 90 nm in the e+ s band,'' in
  \emph{Optical Fiber Communication Conference}.\hskip 1em plus 0.5em minus
  0.4em\relax Optica Publishing Group, 2022, pp. Th4C--8.

\end{thebibliography}
%\begin{thebibliography}{1}
%\end{thebibliography}


\begin{IEEEbiography}[{\includegraphics[width=1in,height=1.25in,clip,keepaspectratio]{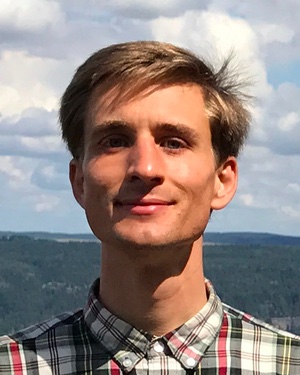}}]{Ryan Hamerly}
was born in San Antonio, Texas in 1988. In 2016 he received a Ph.D. degree in applied physics from Stanford University, California, for work with Prof. Hideo Mabuchi on quantum control, nanophotonics, and nonlinear optics. In 2017 he was at the National Institute of Informatics, Tokyo, Japan, working with Prof. Yoshihisa Yamamoto on quantum annealing and optical computing concepts, and is currently a Senior Scientist at NTT PHI Laboratories and a visiting scientist at MIT, Cambridge, Massachusetts, with Prof.\ Dirk Englund.
\end{IEEEbiography}

\begin{IEEEbiography}[{\includegraphics[width=1in,height=1.25in,clip,keepaspectratio]{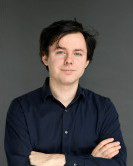}}]{Alexander Sludds}
received his Bachelors of Science and Masters of Engineering in Electrical Engineering and Computer Science from MIT in 2018 and 2019. His research interests focus on the creation of novel CMOS photonic computing and interconnect solutions to enable 1000X improvement over existing commercial technology in the datacenter and edge. He is currently working towards his Ph.D under Prof. Dirk Englund at MIT.
\end{IEEEbiography}

\begin{IEEEbiography}[{\includegraphics[width=1in,height=1.25in,clip,keepaspectratio]{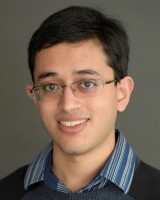}}]{Saumil Bandyopadhyay} received his S.B. in Electrical Engineering and M.Eng. in Electrical Engineering and Computer Science from MIT in 2017 and 2018, respectively. He is a recipient of the NSF Graduate Research Fellowship and is currently a PhD student in Prof. Dirk Englund's group at MIT, where he works on programmable silicon photonics for quantum information processing and artificial intelligence.
\end{IEEEbiography}

\begin{IEEEbiography}[{\includegraphics[width=1in,height=1.25in,clip,keepaspectratio]{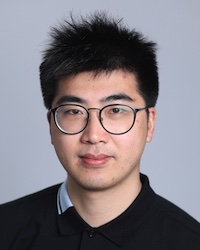}}]{Zaijun Chen} accomplished his Ph.D. degree in Prof. Theodor W. H{\"a}nsch’s group at Max-Planck Institute of Quantum Optics (MPQ) and LMU Munich in 2019, for work on frequency-comb-based precision spectroscopy with Dr. Nathalie Picqu{\'e}.  In 2020, he worked with Prof. Christian Gross on quantum simulation with cold atoms in optical tweezers in Prof. Immanuel Bloch’s group at MPQ.  He moved to MIT as a postdoctoral researcher in Prof. Dirk Englund’s group in 2021, where his research focus is optical computing for machine learning applications. He is currently starting his own research group in the Ming Hsieh Department of Electrical and Computer Engineering at University of Southern California.
\end{IEEEbiography}

%\newpage

\begin{IEEEbiography}[{\includegraphics[width=1in,height=1.25in,clip,keepaspectratio]{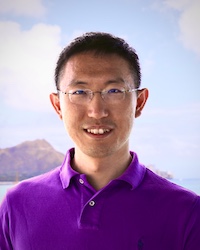}}]{Zhizhen Zhong}
received his Ph.D. and Bachelor degrees in Electronic Engineering from Tsinghua University in 2019 and 2014, respectively. During his graduate studies, he was a visiting Ph.D. student in the Department of Computer Science at the University of California, Davis. Right after finishing the Ph.D., he was a visiting researcher at the network infrastructure team of Meta. Since 2020, he is a postdoctoral researcher working with Prof. Manya Ghobadi at MIT CSAIL. His current research explores the intersection between applied photonics and networked systems to build next-generation computing infrastructures.
\end{IEEEbiography}

\begin{IEEEbiography}[{\includegraphics[width=1in,height=1.25in,clip,keepaspectratio]{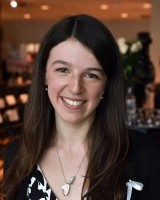}}]{Liane Bernstein}
received her Bachelor of Engineering from Polytechnique Montreal in 2016, specializing in Photonics. There, she worked on Raman spectroscopy and optical coherence tomography for biomedical imaging. In 2018, she earned her Master of Science in Electrical Engineering and Computer Science at MIT for ``Ultrahigh-Resolution, Deep-Penetration Spectral-Domain Optical Coherence Tomography'' in Prof. Andy Yun's group. For her Ph.D. work, she is currently developing both theoretical descriptions as well as experimental demonstrations of optical deep neural networks in Prof. Dirk Englund's group at MIT.
\end{IEEEbiography}

%\begin{IEEEbiography}{Manya Ghobadi}
%needs to add her photo and author biography.
%\end{IEEEbiography}

\begin{IEEEbiography}[{\includegraphics[width=1in,height=1.25in,clip,keepaspectratio]{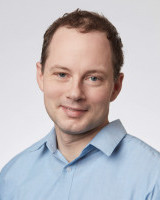}}]{Dirk Englund}
received his BS in Physics from Caltech in 2002. After a Fulbright fellowship at T.U. Eindhoven, he completed an MS in Electrical Engineering and a PhD in Applied Physics at Stanford University in 2008. After a postdoctoral fellowship at Harvard University, he joined Columbia University as Assistant Professor of E.E. and of Applied Physics. He joined the MIT EECS faculty in 2013. Recent recognitions include the 2011 PECASE, the 2011 Sloan Fellowship in Physics, the 2012 DARPA Young Faculty Award, the 2017 ACS Photonics Young Investigator Award, and the OSA's 2017 Adolph Lomb Medal, a Bose Research Fellowship in 2018, and a 2020 Humboldt Research Fellowship.
\end{IEEEbiography}

%\newpage

%\newpage

%\section*{List of TODOs}

%\begin{itemize}
	%\item RIN noise?  Potentially tie to Tait result.  More work though and I'm lazy.
	%\item Fill in the remaining references, unless I'm too lazy.
	%\item Also cite new papers: Alex's paper (duh), SPIE paper, Liane's paper, Ronald's paper, blah blah blah.
	%\item Author biographies.
	%\item Author IEEE / etc. memberships.
	%\item Check LN/S case; I'm not sure that one's correct and consistent with the table.
	%\item Manya wanted to be taken off author list because of not making contributions to this work?
%\end{itemize}

%Congratulations for reading to the very very end!  This page will be cut in the final submission.

%In conclusion, pigs can fly!  And not only can they fly, they can BRRRRRRRRRRRRRRRRRRRRRRRRRRRRRRRRRT. 

%\begin{figure}[htbp]
%\begin{center}
%\includegraphics[width=1.00\columnwidth]{FigTemp.jpeg}
%\caption{Moo moo mooooo moo moo moooo moo mooo mooooo moo moo moo moo mooo moooooooooooooo moo moo moo moo moooo moo moo mooooo moo moo moo moooooo mooooooo mooooo moo moo moo.}
%\label{fig:moo}
%\end{center}
%\end{figure}


%\vfill


\end{document}